\documentclass[aps,pra,twocolumn,superscriptaddress,reprint,longbibliography]{revtex4-1}
\usepackage{amsmath,hyperref,xcolor,amsthm,amssymb,bm}
\usepackage{siunitx}
\usepackage{graphicx}
\usepackage{xcolor}
\usepackage{comment}

\newcommand{\dd}[1]{\mathrm{d}#1\,}

\newcommand{\en}{\epsilon}
\newcommand{\DoSG}{\rho_{\rm G}}
\newcommand{\DoSS}{\rho_{\rm S}}
\newcommand{\DoSGF}{\rho_{{\rm G}0}}
\newcommand{\DoSSF}{\rho_{{\rm S}0}}
\newcommand{\uG}{E_F}
\newcommand{\TG}{T_{\rm G}}
\newcommand{\TGmin}{T_{\rm G,b}}
\newcommand{\TS}{T_{\rm S}}
\newcommand{\TB}{T_{\rm B}}
\newcommand{\Tcross}{T_{\rm G,cr}}
\newcommand{\ThC}{P_{\rm GIS}}
\newcommand{\ThCeph}{P_{\rm e/ph}}

\newcommand{\ThCJ}{P_{\rm J}}
\newcommand{\ThCin}{P_{\rm in}}
\newcommand{\GGIS}{G_{\rm GIS}}
\newcommand{\Geph}{G_{\rm e/ph}}

\newcommand{\RG}{R_{\rm G}}
\newcommand{\Rt}{R_t}
\newcommand{\kB}{k_{\rm B}}

\newcommand{\tauth}{{\tau_{\rm th}}}
\newcommand{\lmfp}{l_{\rm mfp}}
\newcommand{\INoise}{\left\langle I^2\right\rangle}
\newcommand{\PNoise}{\left\langle P^2\right\rangle}
\newcommand{\IPNoise}{\left\langle IP\right\rangle}
\newcommand{\Resp}{{\cal R}}
\newcommand{\NEP}{{\cal N}}
\newcommand{\Vopt}{V_{\rm opt}}
\newcommand{\Mob}{\mu}
\newcommand{\SiO}{$\rm SiO_2\, $}

\begin{document}
	
\title{Electron cooling with graphene-insulator-superconductor tunnel junctions and applications to fast bolometry}

\author{Francesco Vischi}
\email{francesco.vischi@df.unipi.it}
\affiliation{NEST, Istituto Nanoscienze-CNR  and Scuola Normale Superiore, Piazza S. Silvestro 3, I-56127 Pisa, Italy}
\affiliation{Dipartimento di Fisica "E. Fermi", Universit\`{a} di Pisa, Largo Bruno Pontecorvo 3, I-56127 Pisa, Italy}
\author{Matteo Carrega}
\affiliation{NEST, Istituto Nanoscienze-CNR  and Scuola Normale Superiore, Piazza S. Silvestro 3, I-56127 Pisa, Italy}
\author{Alessandro Braggio}
\affiliation{NEST, Istituto Nanoscienze-CNR  and Scuola Normale Superiore, Piazza S. Silvestro 3, I-56127 Pisa, Italy}
\author{Federico Paolucci}
\affiliation{INFN Sezione di Pisa, Largo Bruno Pontecorvo 3, I-56127 Pisa, Italy}
\affiliation{NEST, Istituto Nanoscienze-CNR  and Scuola Normale Superiore, Piazza S. Silvestro 3, I-56127 Pisa, Italy}
\author{Federica Bianco}
\affiliation{NEST, Istituto Nanoscienze-CNR  and Scuola Normale Superiore, Piazza S. Silvestro 3, I-56127 Pisa, Italy}
\affiliation{Dipartimento di Fisica "E. Fermi", Universit\`{a} di Pisa, Largo Bruno Pontecorvo 3, I-56127 Pisa, Italy}
\author{Stefano Roddaro}
\affiliation{Dipartimento di Fisica "E. Fermi", Universit\`{a} di Pisa, Largo Bruno Pontecorvo 3, I-56127 Pisa, Italy}
\affiliation{NEST, Istituto Nanoscienze-CNR  and Scuola Normale Superiore, Piazza S. Silvestro 3, I-56127 Pisa, Italy}
\author{Francesco Giazotto}
\affiliation{NEST, Istituto Nanoscienze-CNR  and Scuola Normale Superiore, Piazza S. Silvestro 3, I-56127 Pisa, Italy}

\begin{abstract}
Electronic cooling in hybrid normal metal-insulator-superconductor junctions is a promising technology for the manipulation of thermal loads in solid state nanosystems. One of the main bottlenecks for efficient electronic cooling is the electron-phonon coupling, as it represents a thermal leakage channel to the phonon bath. Graphene is
a two-dimensional material that exhibits a weaker electron-phonon coupling compared to standard metals.
For this reason, we study the electron cooling in graphene-based systems consisting of a graphene sheet contacted by two insulator/superconductor junctions. We show that, by properly biasing the graphene, its electronic temperature can reach base values lower than those achieved in similar systems based on metallic ultra-thin films. Moreover, the lower electron-phonon coupling is mirrored in a lower heat power pumped into the superconducting leads, thus avoiding their overheating and preserving the cooling mechanisms. Finally, we analyze the possible application of cooled graphene as a bolometric radiation sensor. We study its main figures of merit, i.e., responsivity, noise equivalent power, and response time. In particular, we show that the built-in electron refrigeration allows reaching a responsivity of the order of $\rm 50\, nA/pW$ and a noise equivalent power of order of $\rm 10^{-18}\,  W\, Hz^{-1/2}$  while the response speed is about $\rm 10\,ns$ corresponding to a thermal bandwidth in the order of {\SI{20}{\mega\hertz}}.
\end{abstract}
\maketitle

\section{Introduction}
{Low temperature physics at the micro- and nano-scale has found many practical applications in ultrafast electronics for computing \cite{mukhanov2011,polonsky1993, devoret2013,wallraff2004,vion2002,mooji1999,fornieri2017,paolucci2018logic},{ low-noise high-sensitivity magnetometers \cite{clarke2005,storm2017,giazotto2010,ronzani2014}, radiation sensors, and detectors \cite{koppens2014,du2014,kraus1996,lee2019,tielrooij2015,tielrooij2017,elfatimy2016, guarcello2019,virtanen2018,giazotto2008}}. Hence, finding novel and
efficient cooling schemes is of primary importance \cite{giazotto2006,muhonen2012}.  Typically, ultra-low temperature cryogenics is
accomplished mainly by exploiting $\rm He^3/He^4$ systems consisting of  expensive and bulky machines, with unavoidable issues for space or portable applications. For this reason, important efforts are spent in the field of solid state cooling to to realize micro-refrigerators that can be efficient and scalable to an industrial standard. Many different systems have been proposed, based for example on chiral Hall channels \cite{sanchez2019,giazotto2007,vannucci2015,ronetti2017}, adiabatic magnetization \cite{dolcini2009,manikandan2019}, piezoelectric elements \cite{steeneken2011}, quantum dots \cite{edwards1993,edwards1995,hussein2019}, single ions \cite{rossnagel2016}, and engines based on superconducting circuits \cite{karimi2016,vischi2018,virtanen2017SNS,marchegiani2016,marchegiani2018}.

A cornerstone in this field is the electron refrigeration  in voltage-biased Normal metal-Insulator-Superconductor (NIS) tunnel junctions \cite{nahum1994,leivo1996}. In such a system, the gap of the superconductor acts as an energy filter for the N metal electrons: under an appropriate voltage bias, only the most energetic electrons, i.e., the hottest ones, are able to tunnel into the superconductor, resulting in a decrease of temperature in the N metal   \cite{nahum1994,leivo1996,muhonen2012,giazotto2006}.
The performance of this system is adversely affected by two main phenomena. One consists of an intrinsic thermal leakage owing to the electron-phonon coupling  \cite{giazotto2006}.
The phonons of the metal can be considered as a thermal bath, which temperature is set by the substrate temperature. Phonons interact with
electrons over the metal volume, consequently supplying heat. Secondly, the heat extracted from the N metal warms up the superconducting leads, with the consequent decrease of the superconducting gap and deterioration of the energy filtering over the electrons \cite{courtois2016,nguyen2013, catelani2018}.

{In this paper, we study the graphene refrigeration based on two Graphene-Insulator-Superconductor (GIS) tunnel junctions forming a SIGIS system. Graphene has several interesting properties compared to metals, for example, a charge carrier concentration-dependent density of states \cite{neto2009}, and a weaker and gate tunable electron-phonon coupling \cite{zihlmann2018,chen2012}. The weak electron-phonon coupling arises from the graphene dimensionality \cite{borzenets2013}, as tested in other low-dimension materials \cite{karvonen2007,karvonen2007b}. As a consequence, for given cooling power, a SIGIS can reach lower temperatures respect to a SINIS system.  Moreover, the lower heat current pumped into the leads decreases their adverse heating, making electron cooling more accessible for concrete applications.

A natural application of electron cooling in SIGIS systems concerns the detection of electromagnetic radiation via bolometric effect. It is known that SINIS systems can be used as bolometers, where the built-in refrigeration enhances the responsivity and decreases the Noise Equivalent Power (NEP)   \cite{nahum1993ApplPhysLett,nahum1993IEEE, golubev2001,kuzmin1998,schmidt2005,lemzyakov2018}. A SIGIS-based bolometer inherits the advantages of built-in refrigeration from a SINIS system, combining them with graphene optoelectronic properties \cite{koppens2014}, such as wide energy absorption spectrum, ultra-fast carrier dynamics \cite{dawlaty2008,brida2013,xia2009,efetov2018,mueller2010}, and tunable optical properties via electrostatic doping \cite{li2008,wang2008}. In particular, the lower operating temperature and the weaker electron-phonon coupling allow further decreasing the NEP, while the graphene low heat capacity allows a faster response time compared to a SINIS bolometer.

From the industrial point of view, SIGIS systems may also have high potentiality in wafer-scale integration thanks to the high quality currently reached in large-area graphene production \cite{deng2019}. Moreover, the tunnel junction can be realized with hexagonal Boron Nitride (hBN), which is an insulating material extremely suitable to be combined with graphene due to the crystal similarities. Tunnel barriers based on hBN  represent a valuable alternative to standard metal oxides insulators, simplifying the fabrication into standard steps. \cite{britnell2012}}
		
The paper is organized as follows. Section \ref{sec:model} introduces the device model, the GIS
tunneling, and the thermal model. Section \ref{sec:TGmin} studies the graphene base temperature in a biased SIGIS, also giving a comparison with a standard SINIS system. Section \ref{sec:dynamics} investigates the system response to perturbations and the related dynamical response time. In section \ref{sec:Bolometer}, we study the bolometric properties by focusing on the responsivity and the NEP. {Section \ref{sec:GammaDynes} discusses the impact of the junction quality on the studied properties and yields a quantitative threshold for experiments. Section \ref{sec:comparison} compares our findings with similar bolometric architectures.}  Finally, section \ref{sec:summary} summarizes our main findings.

\section{Model}
\label{sec:model}
\begin{figure}[t]
	\centering
	\includegraphics[width=0.48\textwidth]{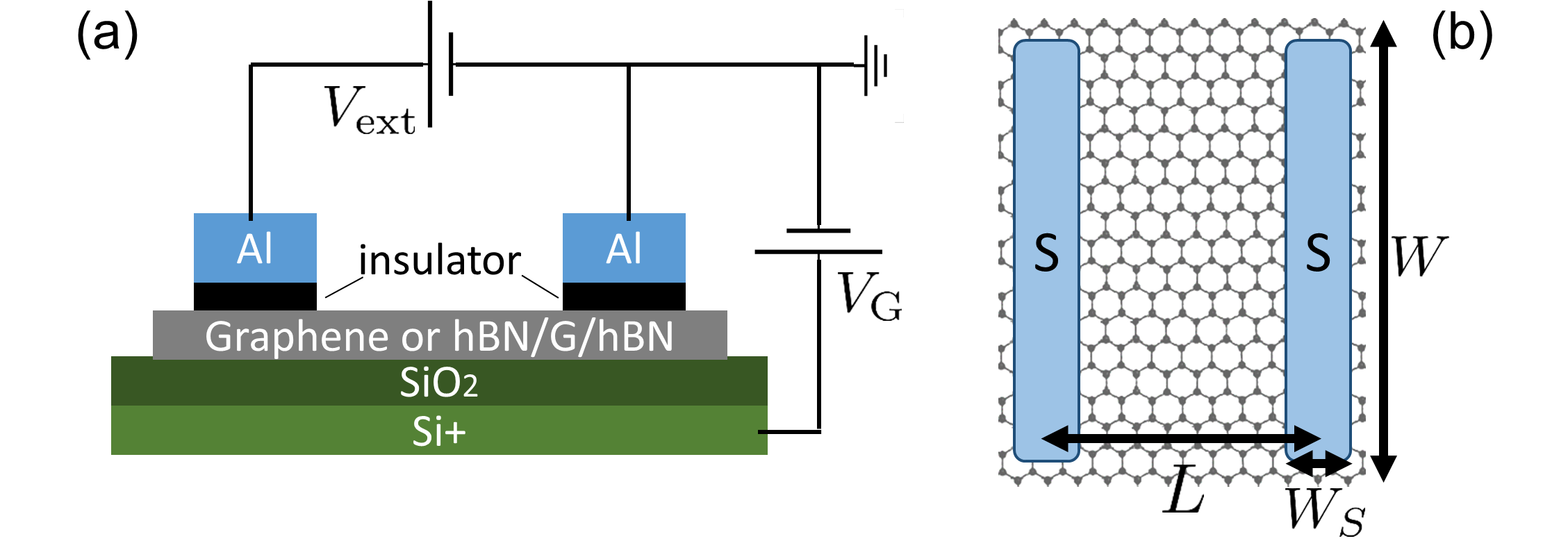}
	\caption{Sketch of the device. (a) A graphene sheet or hBN-encapsulated graphene is in contact with two aluminum leads through a tunnel junction with resistance $\Rt$. In the case of hBN-encapsulated graphene, the insulators can be provided by the hBN layer itself. A voltage bias $V_{\rm ext}$ is applied to the two leads. A back gate, biased with a voltage $V_{\rm G}$, allows tuning carrier density on the graphene sheet. (b) View from the top. The graphene geometrical dimensions are $W$ and $L$, with resulting sheet resistance $\RG$ and area $A$. The superconducting leads are wide $W_S$.
	}
	\label{fig:sketch}
\end{figure}
We consider the system sketched in Fig. \ref{fig:sketch}. It consists of a graphene sheet contacted by two superconducting leads through tunnel junctions of resistance $\Rt$ each. Superconductors are assumed made of aluminum with superconducting gap  $\Delta_0=\SI{200}{\micro\electronvolt}$ and critical temperature $T_c \sim \SI{1.3}{\kelvin}$. The graphene can be deposited directly on \SiO or hBN. The graphene sheet has a rectangular area, with geometrical dimensions $A= W\times L$. The two leads, with dimensions $W\times W_S$, are placed at distance $L$ (see Fig. \ref{fig:sketch}b) and  connected to a voltage generator $V_{\rm ext}$. The electric current $I$  is determined by $\Rt$ and the graphene sheet resistance $\RG= L \rho/W$, where the sheet resistivity $\rho = 1/e n \Mob$ depends on the carrier density $n$ and the electron mobility $\Mob$, being $e$ the modulus of the electron charge. The graphene is gated with a back-gate placed under the substrate and connected to an external generator $V_{\rm G}$.

The proposed setup has many geometrical/fabrication parameters. As a consequence, we fix some of them to reasonable experimental values. By choosing proper geometrical dimensions for the graphene sheet, we consider a negligible sheet resistance compared to the tunnel resistance ($\RG\ll\Rt$). This assumption allows neglecting the voltage partition between the junctions and the sheet. So, the Joule heating of graphene results negligible. To this aim, we set the aspect ratio to $L=W/5$, corresponding to $\RG\approx\SI{250}{\ohm}$ for graphene with $\Mob\approx\SI{5000}{\centi\meter^2/\volt\second}$ and residual carrier density $n_0\approx\SI{1E12}{\centi\meter^{-2}}$, typical for graphene on \SiO \cite{chen2008,martin2007,dassarma2011}. A similar value of resistance can be considered for an encapsulated graphene in a hBN/G/hBN heterostructure, where mobilities are commonly over $\Mob\approx\SI{50000}{\centi\meter^2/\volt\second}$ but the residual charge densities are lower than $n_0\approx\SI{1E11}{\centi\meter^{-2}}$ \cite{amet2014,amet2015,mayorov2012,banszeruse2015}. An advantage of the encapsulated graphene is that the top layer of ultra-thin hBN can be exploited as a high-quality tunnel junction \cite{britnell2012}.

We consider a large graphene area  $A=\SI{100}{\micro\meter^2}$. Large area samples are preferred for bolometric applications since they keep the device in linear response regime and extend the dynamical range of the detector \cite{mckitterick2015,du2014}. Moreover, a greater area reduces the temperature fluctuations, since the thermal inertia due to the heat capacity scales with the area.

Finally, we fix the tunnel resistance as $\Rt=\SI{10}{\kilo\ohm}$. This value is compatible with tunnel junction made of 2-layer hBN \cite{britnell2012,zihlmann2018}  and makes the assumption $\Rt\gg\RG$ valid. We also observe that the tunnel barriers suppress the superconducting proximity effect in graphene.  

{Table \ref{tab:tab1} is a summary of the parameters adopted in the numerical simulations. Some of them will be introduced in the following.}

\begin{table}
	\begin{tabular}{l l}
		\hline
		Graphene dimensions & $L\times W\text=\SI{4.5}{\micro\meter}\times\SI{22}{\micro\meter}$ \\
		Graphene area & $A\text=\SI{100}{\micro\meter^2}$\\
		Residual electron density & $n\text=10^{12}\,{\rm cm}^{-2}$ \\ 
		Tunnel resistance & $\Rt\text=\SI{10}{\kilo\ohm}$ \\
		Sheet resistance & $\RG\text=\SI{250}{\ohm}$ \\
		Electron-phonon coupling & $\Sigma_D\text= \SI{23}{\milli\watt\meter^{-2}\kelvin^{-3}}$ (dirty case) \\
		& $\Sigma_C\text= \SI{24}{\milli\watt\meter^{-2}\kelvin^{-4}}$ (clean case)\\
		Heat capacity at ${\rm 0.5K}$ & $C= \SI{34}{\zepto\joule\kelvin^{-1}}\approx2.4\times10^3\kB$  \\
		\hline
	\end{tabular}
	\caption{Parameters used in the numerical calculations for the device under investigation.}
	\label{tab:tab1}
\end{table}

\subsection{GIS tunneling and cooling}
\label{sec:GIS}
Here, we introduce the main equations and discuss the electron tunneling through a GIS junction. The tunneling rate is proportional to the Density of States (DoS) of graphene and superconductor \cite{maiti2012}. The graphene DoS $\nu_G$ reads \cite{neto2009}
\begin{equation}
\nu_G = \DoSGF \DoSG(\en) \qquad \DoSG(\en) = \frac{\en}{\uG} \, \, ,
\label{eq:GDoS}
\end{equation}
where $\en$ is the energy, $\DoSGF$ is the DoS at the Fermi level, $\DoSG(\en)$ is the normalized graphene DoS and $\uG$ is the Fermi energy. The DoS at the Fermi level is related  to the carrier density by $\DoSGF=2\uG/\pi \hbar^2 v_F^2$ and $\uG=\hbar v_F \sqrt{\pi n}$ where $v_F\approx10^6$ m/s is the Fermi velocity \cite{neto2009} and  $\hbar\approx\SI{6.6E-16}{\electronvolt \cdot \second}$ is the reduced Planck constant. 

The superconductor DoS is 
\begin{align}
&\nu_S= \DoSSF  \DoSS(\en) \, \, ,\\
&\DoSS(\en) = \left| {\rm Re}\frac{(\en+i\Gamma_D)}{\sqrt{(\en+i\Gamma_D)^2-\Delta^2(T)}} \right| \,\,,
\label{eq:SuperconductorDoS}
\end{align}
where $\en$ is the energy, $\DoSSF$ is the DoS at Fermi level of the normal state aluminum, $\DoSS(\en)$ is the superconductor normalized DoS, $\Delta(T)$ is the temperature-dependent superconductivity gap of the Bardeen-Cooper-Schrieffer (BCS) theory { and $\Gamma_D$ is the Dynes parameter that phenomenologically takes into account the subgap tunneling and the smearing of the superconducting peaks, which are also related to the quality of the junction. In this paper, we fixed $\Gamma_D=10^{-4}\Delta_0$, for simplicity. In section \ref{sec:GammaDynes}, we show the dependence of the results on higher values of $\Gamma_D$. }

The charge current in a GIS tunnel junction can be expressed as \cite{maiti2012,zihlmann2018}
\begin{multline}
I(V,\TG,\TS) = \frac{1}{e R_t} \int_{-\infty}^{\infty} \dd \en\left\lbrace \right. \DoSG(\en-eV-E_F) \times \\ 
\left.  \DoSS(\en) [f(\en-eV,\TG)-f(\en,\TS)] \right\rbrace \,\, , 
\label{eq:ElectricalCurrent}
\end{multline}
where $V$ is the voltage drop across the tunnel junction, $\TG$ and $\TS$ are the graphene and superconductor electronic temperatures, respectively. Finally, $f(\en,T)$ is the Fermi distribution. In the following we assume that $V=V_{\rm ext}/2$.

Similarly, the heat current from G to S is
\begin{multline}
\ThC(V,\TG,\TS) = \frac{1}{e^2R_t} \int_{-\infty}^{\infty} \dd \en\left\lbrace 
(\en-eV)\times\right.\\
\left. \DoSG(\en-eV-\uG) \DoSS(\en) [f(\en-eV,\TG)-f(\en,\TS)] \right\rbrace
\label{eq:ThermalCurrent} \, \, .
\end{multline}
We set the sign convention such that $\ThC>0$ means that the heat is extracted from graphene towards superconductors.
It is important to note that when the graphene Fermi energy $\uG$ is much greater than the superconducting gap $\Delta_0$, the graphene DoS dependence on energy can be disregarded in the tunneling integrals, i.e., $\DoSG(\en-eV-\uG)\approx 1$.  Indeed, for $eV,\kB\TS,\kB\TG \lessapprox \Delta_0$, the distribution $[f(\en-eV,\TG)-f(\en,\TS)]$ defines an energy bandwidth of a few $\Delta_0$ around the Fermi level. In this energy window, the graphene DoS  has a variation of the order of $\Delta_0/\uG$ that can be hence neglected when $\uG\gg\Delta_0$. This condition, in general, holds experimentally, as indicated by the presence of a residual charge density $n_0$ \cite{mayorov2012}. The lowest values of residual charge density can be obtained in high quality hBN/G/hBN heterostructures and unlikely goes below $n_0\approx\SI{5E10}{\centi\meter^{-2}}$ \cite{arjmandi_tash2018}; this value corresponds to the lowest value of Fermi energy $\uG=\hbar v_F \sqrt{\pi n_0}\approx \SI{26}{\milli\electronvolt}$, that is at least 100 times the value of $\Delta_0= \SI{0.2}{\milli\electronvolt}$, confirming $\Delta_0\ll \uG$. We remark that the BCS theory provides that $\Delta(T)<\Delta_0$, implying that $\uG\gg \Delta_0>\Delta(T)$, i.e., ensuring that the superconducting gap is lower than the Fermi energy at every temperature.

Therefore, tunneling integrals in Eqs. (\ref{eq:ElectricalCurrent}), (\ref{eq:ThermalCurrent})  take the standard functional form of the NIS tunneling expressions \cite{muhonen2012,giazotto2006,anghel2001,muller1997}. 
We point out that this approximation does not completely drop out the dependence of the tunnel integrals on the Fermi level/carrier density. It is indeed still contained in $\Rt$. We will discuss this point better at the end of this subsection.

\begin{figure}[t]
	\centering
	\includegraphics[width=0.48\textwidth]{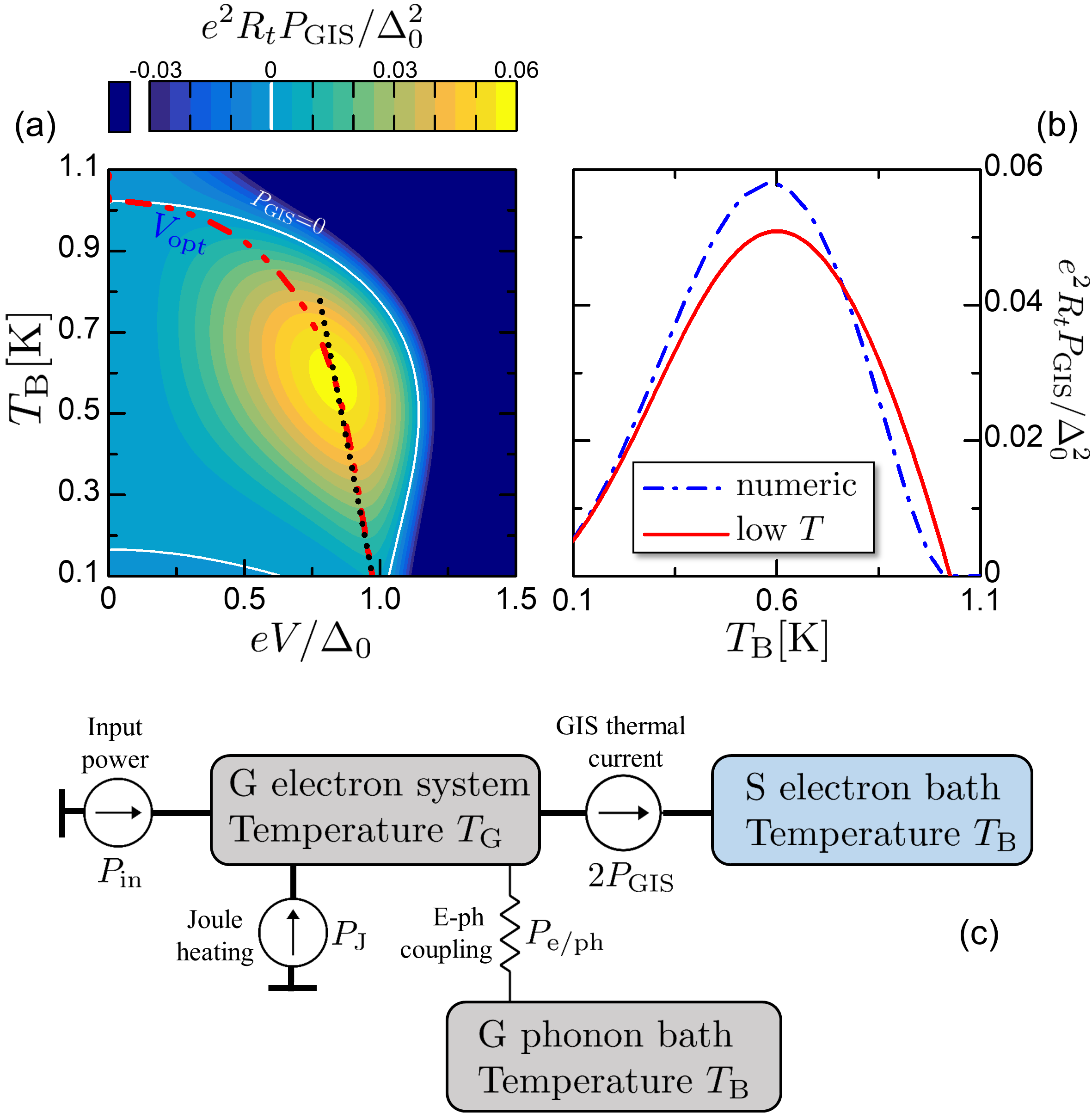}
	\caption{Cooling characteristic of the GIS junction.  (a) Heat current $\ThC$ versus the single junction bias voltage $V$ and the bath temperature $\TB$ when $\TG=\TB$. When $\ThC>0$, the heat is extracted from the graphene. The solid white  line indicates the boundary between the cooling and the heating regions. The red dot-dashed curve represents the optimal bias $\Vopt$ where $\ThC$ is maximized for fixed $\TB$. The black dotted line $eV = \Delta_0- 0.66 \kB \TS$ is the low $\TB$ approximation of $\Vopt$. (b) The value of $\ThC$ at the optimal bias $\Vopt$, obtained by numeric integration of Eq. (\ref{eq:ThermalCurrent}) or by the low $T$ approximation in Eq. (\ref{eq:AnalyticCooling}). (c) Thermal model of a SIGIS. The electron system at temperature $\TG$ is under different heat currents. One is the heat exchange with the graphene phonon bath $\ThCeph$. At the same time, heat is pumped away by the junction with power $2\ThC$ and released in the superconductor electron bath. Another source of heat is the Joule heating given by the flowing electric current.
	}
	\label{fig:CoolingMap}
\end{figure}
Figure \ref{fig:CoolingMap}a displays the behavior of $\ThC$ versus $V$ and $\TG$ is equal to the bath temperature $\TB$. In the regions where $\ThC>0$, the heat is extracted from graphene, implying electron cooling. It corresponds to the yellow-green area delimited by the white curve ($\ThC=0$). The cooling power is maximized, for given value of $\TB$, at the optimal voltage bias  $\Vopt(\TB)$ (see red curve in Fig. \ref{fig:CoolingMap}a). The cooling power value along the $\Vopt$ curve is reported in Fig. \ref{fig:CoolingMap}b as function of $\TB$. The maximum is about $\ThC\approx0.06\Delta_0^2/(e^2 \Rt)$  for $\TB\approx0.6\,$K$\,\approx T_c/2$ and $V\approx0.82\Delta_0/e$ ($\sim\SI{170}{\micro\volt}$ for aluminum). For $\Rt=\SI{10}{\kilo\ohm}$, the maximum cooling power corresponds to about $\ThC\approx\SI{0.24}{\pico\watt}$. 

Low temperature ($\TS,\TG\ll \Delta_0/\kB$) approximated expressions of Eqs. (\ref{eq:ElectricalCurrent}) and (\ref{eq:ThermalCurrent}) are reported in Refs.  \cite{giazotto2006,muhonen2012,anghel2001,muller1997}. In this approximation, the optimal cooling is $e\Vopt \approx \Delta_0- 0.66 \kB \TS$ (see the dotted black curve in Fig. \ref{fig:CoolingMap}a), corresponding to an electric current
\begin{equation}
I \approx 0.48 \frac{\Delta_0}{e \Rt} \sqrt{\frac{\kB \TG}{\Delta_0}} \, \, ,
\label{eq:AnalyticCurrent}
\end{equation}
and a related cooling power
\begin{equation}
\ThC \approx \frac{\Delta_0^2}{e^2 \Rt}\left[ 0.59 \left(\frac{\kB \TG}{\Delta_0}\right)^{3/2} - \sqrt{\frac{2\pi \kB \TS}{\Delta_0}}e^{-\Delta_0/\kB \TS}
\right] \, \, .
\label{eq:AnalyticCooling}
\end{equation}

Before concluding this section, we wish to discuss the dependence of equations (\ref{eq:ElectricalCurrent}) and (\ref{eq:ThermalCurrent}) on the carrier density $n$ and how this can affect the electronic and thermal transport. The carrier density $n$ is tuned via field effect by the gate voltage $V_{\rm G}$ (see Fig. \ref{fig:sketch}a). The electric and thermal currents depend on $n$ through the tunnel resistance $\Rt$. The latter is proportional to the DoS of both graphene and superconductor and to the modulus square of the tunneling amplitude  $|U_0|^2$, i.e. $\Rt\propto1/(\DoSGF \DoSSF |U_0|^2)$ \cite{maiti2012,heikkilabook}. Since $\DoSGF\propto\sqrt{n}$, the GIS tunnel resistance depends on the carrier density as
\begin{equation}
\Rt(n)  = \Rt(n\text = n_0) \sqrt{\frac{n_0}{n}} \, \, ,
\label{eq:RtOnn}
\end{equation}
where $n_0$ is the residual carrier density. This equation implies
\begin{align} 
&I(V,\TG,\TS,n)  = I(V,\TG,\TS,n=n_0) \sqrt{\frac{n}{n_0}} \\
& \ThC(V,\TG,\TS,n)  = \ThC(V,\TG,\TS,n=n_0) \sqrt{\frac{n}{n_0}} \, \, .
\label{eq:CurrentsOnn}
\end{align}
This simple scaling on $n$ is valid when the approximation $\rho(\en-eV-\uG)\approx 1$, i.e. when $\uG\gg \Delta_0$. This condition is experimentally respected since charge density $n$ can be tuned typically in a range from $\SI{5E10}{\centi\meter^{-2}}$ to $\SI{5E13}{\centi\meter^{-2}}$, when using standard solid gating. This range is experimentally limited on the bottom by the presence of charge puddles  \cite{martin2007} and on the top by the occurrence of gate dielectric breakdown caused by high voltage.

\subsection{SIGIS Thermal model}
\label{sec:thermalmodel}
In this section, we describe the thermal model that includes all the thermal channels to graphene, as sketched in Fig. \ref{fig:CoolingMap}c. We consider the graphene sheet homogeneously at the same temperature, neglecting the spatial dependence of $\TG$, thanks to the high heat diffusivity in graphene \cite{dawlaty2008,brida2013,xia2009}. Moreover, we treat the graphene phonon bath as a reservoir at a fixed temperature $\TB$. This assumption is physically reasonable owing to the negligible Kapitza thermal resistance between the graphene and the substrate \cite{balandin2011,chen2009}. Finally, we consider the superconductor electrons as a thermal reservoir well thermalized with the substrate, by imposing $\TS=\TB$. This assumption can be violated in real experiments, where the heat pumped into the superconductor heats up its quasi-particles, and the weak electron/phonon (e/ph) coupling provides a poor cooling to the bath \cite{giazotto2006}. This effect is detrimental for the superconducting state and, as a consequence, for cooling. In general, this effect can be weakened by contacting the superconductor with hot quasi-particles traps or coolers in cascade \cite{courtois2016, nguyen2013, catelani2018, nguyen2016,camarasa2014}, making our assumption physically reasonable. Moreover, in a SIGIS system, the amount of heat transferred into the superconductor is lower than that present in a SINIS system, because of the lower heat leakage from the phonon bath to the graphene electrons. 

Thus, in our thermal model (see Fig. \ref{fig:CoolingMap}c) the only variable is the graphene temperature $\TG$, which is determined by the solution of the following heat balance equation
\begin{multline}
C(\TG) \frac{\dd \TG }{\dd t} + 2\ThC(\TG,\TB,V)+\\ +\ThCeph(\TG,\TB)-\ThCJ(\TG,\TB,V)= \ThCin  \, \, .
\label{eq:BalanceEquation}
\end{multline}
This equation takes into account the heat current across the two junctions $2\ThC$, the electron-phonon coupling in graphene $\ThCeph$, the Joule heating $\ThCJ$ and a possible external power input $\ThCin$ (for example a radiation power) that we consider to investigate the bolometric response. We also consider the time dependence of $\TG$ introducing the electron heat capacity $C$, which plays the role of thermal inertia of the system when dynamic response is investigated. 

Let us consider the electron-phonon heat current $\ThCeph$. Below the Bloch-Gr\"{u}neisen temperature ($\rm\sim50\,K$), $\ThCeph$ is characterized by the presence of two different regimes depending on whether the wavelength of thermal phonons is longer or shorter than the electron mean free path $\lmfp$ \cite{zihlmann2018,chen2012,efetov2010,hwang2008,gonnelli2015}. In the clean regime (or short wavelength regime) the e/ph coupling reads
\begin{align}
& \ThCeph = A \Sigma_C \left(\TG^4-\TB^4\right)\\
&\Sigma_C = \frac{\pi^{5/2} D_p^2 \sqrt{n} \kB^4}{15 \rho_M \hbar^4 v_F^2 s^3} \, \, ,
\label{eq:PhononsClean}
\end{align}
while in the dirty regime (or long wavelength regime) takes the form
\begin{align}
& \ThCeph = A \Sigma_D \left(\TG^3-\TB^3\right)\\
&\Sigma_D = \frac{2 \zeta(3) D_p^2 \sqrt{n} k_B^3}{\pi^{3/2} \rho_M \hbar^3 v_F^2 s^2 \lmfp} \, \, , 
\label{eq:PhononsDirty}
\end{align}
where $\Sigma_C$, $\Sigma_D$ are the electron-phonon coupling constants, depending on the sound speed $s \approx \SI{2E4}{\meter/\second}$, the mass density $\rho_M \approx \SI{7.6E-7}{\kilo\gram/\meter^2}$, the deformation potential $D_p \approx \SI{13}{\electronvolt}$, $\lmfp\approx\SI{60}{\nano\meter}$ and the Riemann Zeta $\zeta(3)\approx 1.2$. As final result, the coupling constants are $\Sigma_C\approx \SI{0.024}{\pico\watt\micro\meter^{-2}\kelvin^{-4}}$ and $\Sigma_D\approx \SI{0.023}{\pico\watt\micro\meter^{-2}\kelvin^{-3}}$ \cite{walsh2017,zihlmann2018,chen2012,viljas2010,neto2009,mckitterick2015,paolucci2017}.

In the following we consider both the graphene regimes, writing the generic coupling $\ThCeph=A \Sigma_\delta (\TG^\delta-\TB^\delta)$, where $\delta$ can be 3 or 4 according  to a dirty or clean regime respectively and $\Sigma_{\delta}$ is $\Sigma_C$ or $\Sigma_D$ coherently. In the temperature range between $\rm0.1\,K$ to $\rm 1\,K$, graphene on \SiO shows a dirty regime, while the hBN-encapsulated graphene is in a clean regime  \cite{mckitterick2015,zihlmann2018}. The reason is the different mobility (and therefore different electron mean free path) due to the presence of the hBN-encapsulation \cite{chen2012,mckitterick2015,zihlmann2018}.

The effect of the two regimes can be evaluated by the electron-phonon thermal conductance $\Geph$ in a system where $\TG$ is perturbed from the equilibrium. $\Geph$ is calculated by the linear expansion $\ThCeph\approx \Geph (\TG-\TB)$ where
\begin{equation}
\Geph=\left.\frac{\partial\ThCeph}{\partial \TG}\right|_{\TB}=
\left\lbrace 
\begin{array}{l}
3 A\Sigma_D \TB ^2,\, \,  {\rm dirty\,regime} \\  4 A\Sigma_C \TB ^3,\, \,   {\rm clean\,regime}
\end{array}
\right.  \, \, .
\label{eq:Geph1}
\end{equation} 
The $\Geph$ in the two regimes are of the same order of magnitude at $\TB=1\,$K, but the different temperature scaling makes the clean regime weaker compared to the dirty one when $\TB$ is below $\rm 1\,K$.

The Joule heating is due to the electron current flow in the resistive sheet of graphene. It is given by $\ThCJ= \RG I^2(\TG,\TB,V)$ and is a component that spoils cooling. In this system, the current-voltage characteristic is non-linear, and the current is suppressed by the presence of the superconductor gap. The Joule heating scales as $\sim\Delta_0^2 \RG/(e\Rt)^2$, while the cooling power as $\sim \Delta_0^2/e^2\Rt$. The ratio between the Joule heating and the cooling power then scales as $\sim\RG/\Rt$, implying that the cooling performance is not affected by the Joule effect when $\RG\ll \Rt$. Indeed, we found out in our simulations that Joule heating weakly affects the thermal equilibrium, which is instead dominated by the competition between $\ThC$, $\ThCeph$, and $\ThCin$. For this reason, we neglect the Joule heating in the analytic results, while we keep it in the numerical ones.

{We remark that, in our thermal model, we do not include the photonic and the phenomenological back-tunneling channels \cite{walsh2017,schmidt2004, meschke2006,muhonen2012, jochum1998}. These two contributions are indeed dependent on the fabrication parameters, such as the device design and on the junction quality. For this reason, they are often considered as empirical parameters to fit the experimental data. Moreover, in the range of temperatures studied in this paper (above 0.1 K), the photonic thermal conductance in our device is negligible compared to the phononic thermal conductance \cite{walsh2017}. Finally, the quasi-particle back-scattering can be managed by adjusting the tunnel resistance of the junction.}

The heat capacity for $\kB \TG\ll \uG$ is given by the standard Fermi liquid result \cite{falkovsky2013, viljas2010, walsh2017}
\begin{equation}
C = A \gamma T \,\, ,
\end{equation}
where $\gamma=(\pi^2/3) \kB^2 \rho_{{\rm G}0}$ is the Sommerfeld coefficient. We notice that the linear behavior of $C$ in temperature owes to the fact that $\kB \TG\ll\uG$, yielding the same behavior of a metal. The dependence of $C$ on the Fermi energy (and hence by the residual charge by $\uG=\hbar v_F \sqrt{\pi n}$) is contained in $\gamma\propto\rho_{{\rm G}0}(\uG) \propto \sqrt{n} $. 

Finally, we comment on the dependence of the heat current contributions on carrier density. {For simplicity, we assume a homogeneous charge density $n$ over the whole graphene area, even though under the metallic contacts the screening may slightly affect this assumption. Anyway, since cooling require very small potential differences ($\approx1\,$mV) between the contacts and graphene, the electron density under the electrodes can be considered constant.
Hence, the carrier density of the whole graphene sheet can be tuned mainly by the backgate, with negligible charge inhomogeneities due to the  specific electrostatic problem. We recall that the sheet resistivity is given by $\rho \propto 1/n$, implying 
\begin{equation}
\RG(n)  = \RG(n\text = n_0) \frac{n_0}{n} \, \, .
\end{equation}
This equation and $\Rt(n)$ in Eq. (\ref{eq:RtOnn}) return that $\ThCJ\propto\RG/\Rt^2$ does not depend on $n$. Moreover, considering Eq. (\ref{eq:CurrentsOnn}) and $\ThCeph\propto \sqrt{n}$, the heat balance equation can be written as
\begin{multline}
2\sqrt{\frac{n}{n_0}}\ThC(\TG,\TB,V,n\text=n_0)+\\ +\sqrt{\frac{n}{n_0}}\ThCeph(\TG,\TB,n\text=n_0)-\ThCJ(\TG,\TB,V,n\text=n_0)=\\= \ThCin - \sqrt{\frac{n}{n_0}} C(\TG,n\text=n_0) \frac{\dd \TG }{\dd t} \, \, .
\label{eq:BalanceEquationOnDensity}
\end{multline}
The dominant terms $\ThC$ and $\ThCeph$ scale as $\sqrt{n}$. The terms that are constant in $n$ are the Joule heating and the external power input $\ThCin$. Hence, the thermal properties are weakly affected by the graphene carrier density if Joule heating is negligible and $\ThCin=0$. The heat balance equation in presence of an external source ($\ThCin\neq0$) will be discussed in section \ref{sec:Bolometer}.

\section{Base temperature}
\label{sec:TGmin}
\begin{figure}[t]
	\centering
	\includegraphics[width=0.48\textwidth]{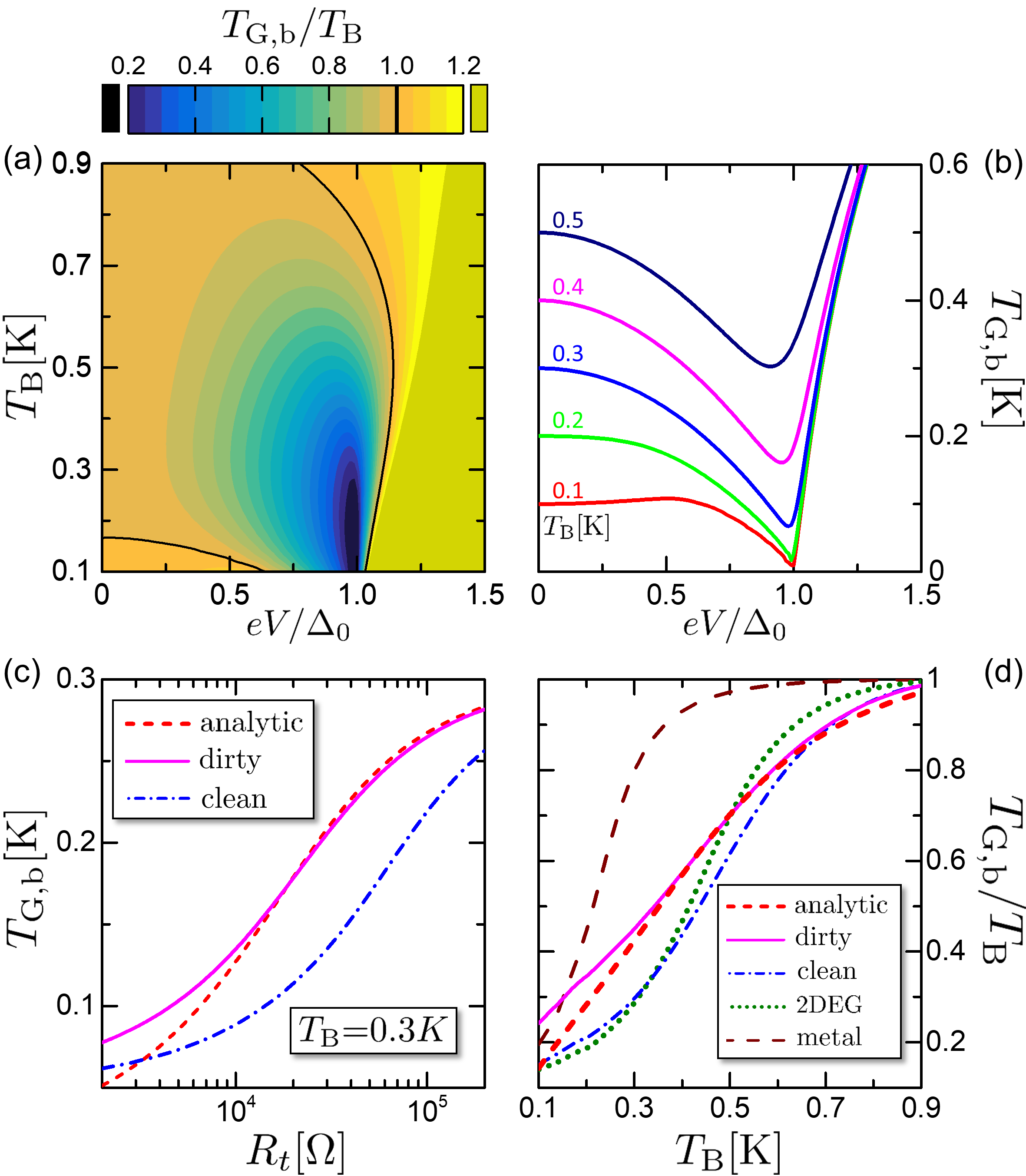}
	\caption{(a)  Color map of the ratio of the base temperature $\TGmin$ with the bath temperature $\TB$ versus $V$ and $\TB$ in the clean electron-phonon regime. The black line shows $\TGmin/\TB=1$ (no cooling) and delimits the region where cooling is present. (b) $\TGmin$ versus $V$ for different bath temperatures $\TB$. (c) $\TGmin$ versus tunnel resistance $\Rt$ for fixed $\TB=0.3\,$K. The analytic curve plots Eq. (\ref{eq:AnalyticTG}). The other curves are calculated numerically for the case of graphene in dirty and clean regime. (d) Comparison of $\TGmin/\TB$ in different materials, for the same area $A$ and tunnel resistance $\Rt$. We plot the results for graphene in dirty and clean regime, the analytical result in Eq. (\ref{eq:AnalyticTG}) for dirty graphene, an ultra-thin metal film of thickness $\rm 1\,nm$ and an InGaAs 2DEG.
	}
	\label{fig:CoolingSimulation}
\end{figure}

In this section, we investigate the stationary ($\partial_t \TG=0$) quasi-equilibrium case of the heat balance equation (\ref{eq:BalanceEquation}) in the absence of external input power ($\ThCin=0$). Solving the balance equation for $\TG$, we can calculate the base temperature $\TGmin$ reached by cooled graphene.

Fig. \ref{fig:CoolingSimulation}a reports a color map of $\TGmin/\TB$ versus $(V,\TB)$ for the case of clean graphene regime. The black line for $\TG/\TB=1$ separates the region of cooling and heating of graphene. Figure \ref{fig:CoolingSimulation}b reports $\TG$ versus $V$ for chosen values of bath temperature $\TB$. When $V\to 0$, the graphene temperature tends to the equilibrium with the bath temperature $\TB$. The minimum temperature is reached when the voltage bias is set closely below $\Delta(T)/e$. In the dirty regime, the cooling behavior is qualitatively similar but lower in performance compared to that in the clean graphene regime, due to the stronger e/ph thermal conductance (see Eq. (\ref{eq:Geph1})), implying higher base temperatures.

When Joule effect is negligible, the base temperature is given by the equilibrium between the electron-phonon heating power and the junction cooling power. The former scales as the area $A$, while the latter scales as $\ThC\propto \Rt^{-1}$. As a consequence, the base temperature is lowered  by decreasing the factor $A \Rt$. The junction resistance cannot be decreased at will since the $\RG \ll \Rt$ condition must be satisfied; otherwise, the detrimental Joule heating contribution is not negligible, and the voltage partition between sheet and junctions must be properly considered. 

The heat balance equation can be solved analytically at optimal bias and low temperatures if the Joule heating is negligible and if the graphene is in the dirty regime. With these assumptions, Eq. (\ref{eq:AnalyticCooling}) can be used for $\ThC$ and then the heat balance equation has a polynomial form that can be solved exactly. On the opposite, the $\TG^4$ form of the e/ph coupling in clean regime yields a not analytically solvable balance equation. The analytic solution is obtained by substituting $\ThC$ with the Eq. (\ref{eq:AnalyticCooling}) and $\ThCeph$ with Eq. (\ref{eq:PhononsDirty}) in the thermal balance equation $2 \ThC  + \ThCeph=0$, yielding
\begin{multline}
\frac{2 \Delta_0^2}{e^2 \Rt}\left[ 0.59 \left(\frac{\kB \TG}{\Delta_0}\right)^{3/2} - \sqrt{\frac{2\pi \kB \TB}{\Delta_0}}e^{-\Delta_0/\kB \TB}
\right] +\\
+ A\Sigma_D(\TG^3-\TB^3) = 0 \, \, ,
\end{multline}
that is a second-order equation $y^2+2by-c=0$ in $y=(\kB \TG/\Delta_0)^{3/2}$ and 
\begin{align}
\nonumber
b = &\frac{0.59\kB^3}{A\Sigma_D\Delta_0 e^2 \Rt}\\
\nonumber
c = & \left(\frac{\kB\TB}{\Delta_0}\right)^3 + \frac{2\kB^3}{A\Sigma_D \Delta_0 e^2 \Rt} \sqrt{\frac{2\pi \kB \TB}{\Delta_0}}e^{-\Delta_0/\kB \TB} \,\,,
\end{align}
with physical solution
\begin{equation}
\label{eq:AnalyticTG}
\TGmin = \frac{\Delta_0}{\kB} \left(\sqrt{b^2+c} -b\right)^{2/3} \, \, .
\end{equation}

Fig. \ref{fig:CoolingSimulation}c reports the dependence of $\TGmin$ on $\Rt$ calculated numerically in case of dirty and clean regimes. The analytical result of Eq. (\ref{eq:AnalyticTG}) for $\TGmin$ in the dirty regime is represented by the red dashed line. We can notice that decreasing $\Rt$ further reduces the achievable base temperature. The agreement between the numeric and analytic results for $\TGmin$ in the dirty regime is generally good if $\TGmin/\TB\approx1$. When $\TGmin/\TB \ll 1$, the solution depends on the accuracy of the $\ThC$ approximation with the consequence that the leading order approximation of $\ThC$ in Eq. (\ref{eq:AnalyticCooling}) is not anymore sufficient. 

In order to investigate the advantage of graphene e/ph coupling, we make a comparison of the base graphene temperature in a SIGIS with the base temperature of a tunnel-cooled system based on a metallic thin film and a two-dimensional electron gas (2DEG). To this aim, we solve the balance equation $2\ThC+\tilde{P}_{\rm e/ph}=0$ for the different systems, where $\ThC$ is the same but $\tilde{P}_{\rm e/ph}$ is  the electron-phonon heat current in a metallic thin film or in a conventional 2DEG with parabolic band dispersion \cite{giazotto2006PRL}. For simplicity, we neglect the resistances of metal and 2DEG and the related Joule heating. For the sake of comparison, we consider the same $A$ and $\Rt$. For a metallic thin film, it is $\tilde {P}_{\rm e/ph}=A w \Sigma_{\rm N}(T_{\rm e}^5-\TB^5)$ and $\Sigma_{\rm N}=\SI{1}{\nano\watt\micro\meter^{-3}\kelvin^{-5}}$, where $T_e$ is the electron temperature. We consider a low thickness $w=1\,$nm, for which we have a coupling per unit area $w \Sigma_{\rm N}\approx\SI{1}{\pico\watt\micro\meter^{-2}\kelvin^{-5}}$. For a 2DEG in $\rm In_{0.75}Ga_{0.25}As$, we have $\tilde{P}_{\rm e/ph}=A  \Sigma_{\rm 2DEG}(T_{\rm e}^5-\TB^5)$ and a coupling per unit area $\Sigma_{\rm 2DEG}\approx\SI{0.073}{\pico\watt\micro\meter^{-2}\kelvin^{-5}}$ \cite{giazotto2006PRL,price1982,ma1991}. At a temperature of the order of 1K, the coupling per unit area of the metal is about 40 times larger than that of graphene, while the coupling per unit area of the 2DEG is about 3 times larger. It can be expected that graphene and 2DEG can reach lower temperatures compared to the metallic thin film. This is shown in Fig. \ref{fig:CoolingSimulation}d, reporting the base temperatures of the different systems.

Deeper insight can be reached by comparing the e/ph thermal conductance per unit area of the different systems. We have in a metal $G_{\rm N}/A=5 w \Sigma_{\rm N} \TB^4$, in a 2DEG $G_{\rm 2DEG}/A=5\Sigma_{\rm 2DEG}\TB^4$ and in graphene $\Geph/A=\delta\Sigma_\delta \TB^{\delta-1}$, with $\delta$ indicating different e/ph regime. It can be noticed that the former two have a better scaling behavior compared to graphene. However, in metals, the coupling constant is large enough that this advantage is effective only below $\TB=0.1\,$K, i.e., below the typical temperature range for the tunnel cooling. This can be seen in Fig. \ref{fig:CoolingSimulation}d where the metal curve reaches the graphene curves (dirty and clean) at about $\rm 0.1\,K$. We remark that a $\rm1\,nm$ thick metallic film is very challenging to be produced.  A different conclusion holds for the 2DEG where the coupling constant $\Sigma_{\rm 2DEG}$ is low enough that the $T^5$ scaling of $\tilde{P}_{\rm e/ph}$ can allow for a lower e/ph heat current in the temperature interval of interest. This can be seen in Fig. \ref{fig:CoolingSimulation}d, where the 2DEG reaches the base temperature of graphene  at $T\approx0.5\,$K for dirty regime and at $T=0.3\,$K for clean regime. This indicates that cooling performances for a 2DEG and a SIGIS are comparable. In this case, the main (and non-trivial) advantage in graphene relies on the fabrication issues. Indeed, the growth of III-IV materials for 2DEGs requires molecular beam epitaxy that is an expensive technique. Furthermore, the use of 2DEGs implies the use of several steps of lithography, etching, and evaporation of metals. On the opposite, Chemical Vapor Deposition is nowadays an established and cheaper technique for growing graphene or hBN/graphene/hBN heterostructures \cite{banszeruse2015}, allowing easier scalability to industrial standards.

\section{Thermal Response Dynamics}
\label{sec:dynamics}
In this section,  we study the dynamics of the SIGIS with thermal perturbations from the base temperature, focusing on its response time. The latter is an important parameter for any time-dependent application since it affects the thermal bandwidth of the system. 

The response time is a parameter that appears in the transfer functions and involves thermal properties, such as the power-to-temperature transfer function or the bolometric responsivity. Both these quantities are studied below.

\begin{figure}[t]
	\centering
	\includegraphics[width=0.48\textwidth]{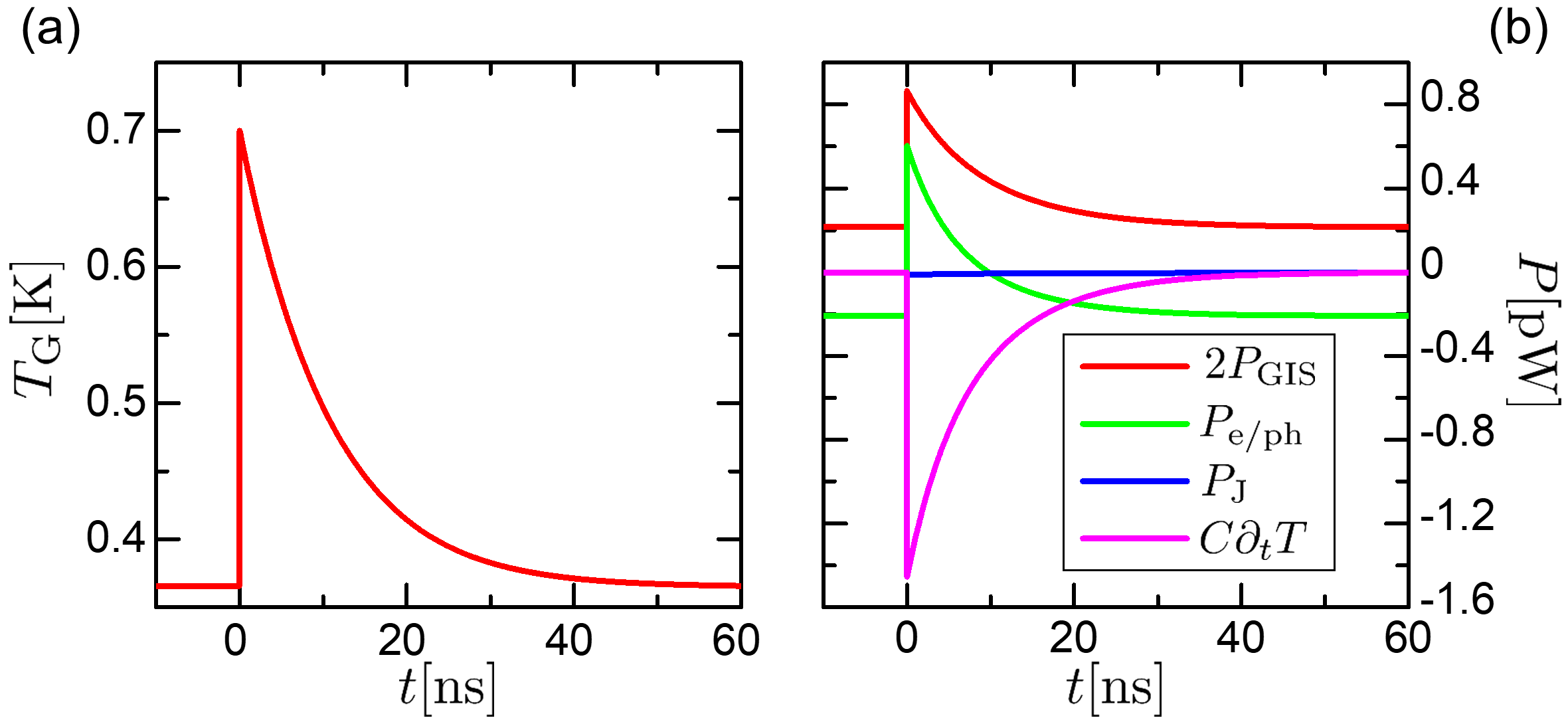}
	\caption{(a) Time evolution of $\TG$ after a power pulse that brings the graphene from $\TGmin\approx0.37\,$K to $\TG=0.7\,$K. Here the bath temperature is  $\TB=0.5\,$K, $V$ is at the optimal bias  $eV\approx0.87\Delta_0$ and the graphene is in dirty regime. (b) Time evolution of the heat current components in Eq. (\ref{eq:BalanceEquation}) corresponding to temperature in panel (a). The Joule component $\ThCJ$ is negligible compared to junction and electron-phonon components. At equilibrium, $2\ThC\approx -\ThCeph$. 
	}
	\label{fig:tauExample}
\end{figure}
As an example of thermal response, we report in Fig. \ref{fig:tauExample} the numerical solution of the heat balance equation (\ref{eq:BalanceEquation}) at bath temperature $\TB=0.5\,$K, optimal voltage bias $e \Vopt(\TB)\approx0.87\Delta_0$ and dirty graphene regime. Figure \ref{fig:tauExample}a shows the evolution of temperature over time. At $t<0$, the graphene is at base temperature $\TGmin\approx 0.37\,$K. The input power is null for the whole process, except at $t=0$, where a power pulse drives the graphene temperature from $\TGmin$ to $\TG=0.7\,$K. After this pulse, the graphene thermalizes to the bath temperature in about 50 ns. The associated heat currents evolution is plotted in Fig. \ref{fig:tauExample}b. In the whole process, it is $2\ThC+\ThCeph+C(\TG)\partial_t \TG=0$. At $t<0$, the graphene is in a stationary state, where $\partial_t \TG = 0$ and the equilibrium is given by $2\ThC+\ThCeph=0$. From Fig. \ref{fig:tauExample}b it can be noticed that the numerical calculations yield an always negligible Joule heating.

Important physical insight into the dynamics can be obtained by studying small perturbations from base temperature by linearizing the heat balance equation. Therefore, we consider the left hand side of Eq. (\ref{eq:BalanceEquation}) in a series expansion around $\TG=\TGmin$ and we assume a constant heat capacity for small perturbations: $C(T)\approx C(\TGmin)$. Moreover, we neglect Joule heating. In this way, we have the linearized thermal equation
\begin{equation}
C \frac{\dd \Delta \TG }{\dd t} + (2\GGIS + \Geph) \Delta \TG = 0  \, \, ,
\label{eq:LinearBalanceEquation}
\end{equation}
where $\Delta \TG = \TG-\TGmin$, and $\GGIS$ and $\Geph$ are thermal conductances related to the junction and the e/ph coupling, respectively. The first term is 
\begin{multline}
\GGIS= \left.\frac{\partial\ThC}{\partial \TG}\right|_{\TGmin} =\\ = \frac{1}{e^2R_t} \int_{-\infty}^{\infty} \dd \en\left\lbrace 
\frac{(\en-eV)^2}{4 \kB \TGmin^2} \frac{1}{\cosh^2\left(\frac{\en-eV}{2\kB \TGmin}\right)}\times \right.\\ \left. \DoSG(\en-eV-\uG) \DoSS(\en)  \right\rbrace \approx\\ \approx \frac{3\cdot 0.59}{2}  \frac{\Delta_0 \kB}{e^2 \Rt} \left(
\frac{\kB \TGmin}{\Delta_0} 
\right)^{1/2} \,\, ,
\label{eq:GGISBiased}
\end{multline}
where the approximation in the last passage is valid at $\Vopt$ and $\TB,\TG\ll\Delta_0/\kB$. The e/ph channel $\Geph$ is given by Eq. (\ref{eq:Geph1}) evaluated at the equilibrium point $\TG=\TGmin$. 

The solutions of the linearized thermal balance equation (\ref{eq:LinearBalanceEquation}) have the exponential form $\Delta \TG \propto e^{- t/\tauth}$, where $\tauth$ is the response time at $\Vopt$ given by
\begin{equation}
\tauth = \frac{C}{G_{\rm tot}} \approx \frac{A \gamma \TGmin}{\delta A \Sigma_\delta \TGmin^{\delta-1} + 1.8 \frac{\Delta_0 \kB}{e^2 \Rt} \left(
	\frac{\kB \TGmin}{\Delta_0} 
	\right)^{1/2} } \, \, .
\label{eq:TimeConstantAnalytics}
\end{equation}

The denominator in Eq. (\ref{eq:TimeConstantAnalytics}) is the sum of the  junction and e/ph thermal conductances. The different temperature scaling of $\GGIS$ and $\Geph$ implies two regimes defined by the dominance of one of the two channels. The two regimes are separated by a crossover temperature $\Tcross$ that can be estimated by equation $\GGIS(\TGmin)=\Geph(\TGmin)$, yielding:
\begin{equation}
\Tcross = \left(\frac{1.8 \Delta_0^{1/2} \kB^{3/2}}{e^2 \Rt \delta A \Sigma_\delta}\right)^{1/(\delta-1.5)}.
\label{eq:TCrossover}
\end{equation}
We obtain $\Tcross=0.39\,$K for dirty graphene  regime and $\Tcross=0.53\,$K for clean graphene  regime. When $\TGmin\ll\Tcross$ the junction conductance dominates over the e/ph conductance and $\tauth$ is
\begin{equation}
\tauth \sim \frac{A\gamma e^2 \Rt }{1.8 \kB^2 } \left(\frac{\kB \TGmin}{\Delta_0}\right)^{1/2}\, \, .
\label{eq:LowT}
\end{equation}
For $\TGmin\gg \Tcross$, there is a regime dominated by the e/ph coupling, yielding
\begin{equation}
\tauth \sim \frac{\gamma \TGmin^{2-\delta}}{\delta \Sigma_{\delta} } \, \, ,
\label{eq:HighT}
\end{equation}
that depends only on the graphene properties and not on geometrical parameters of the SIGIS.

\begin{figure}[t]
	\centering
	\includegraphics[width=0.48\textwidth]{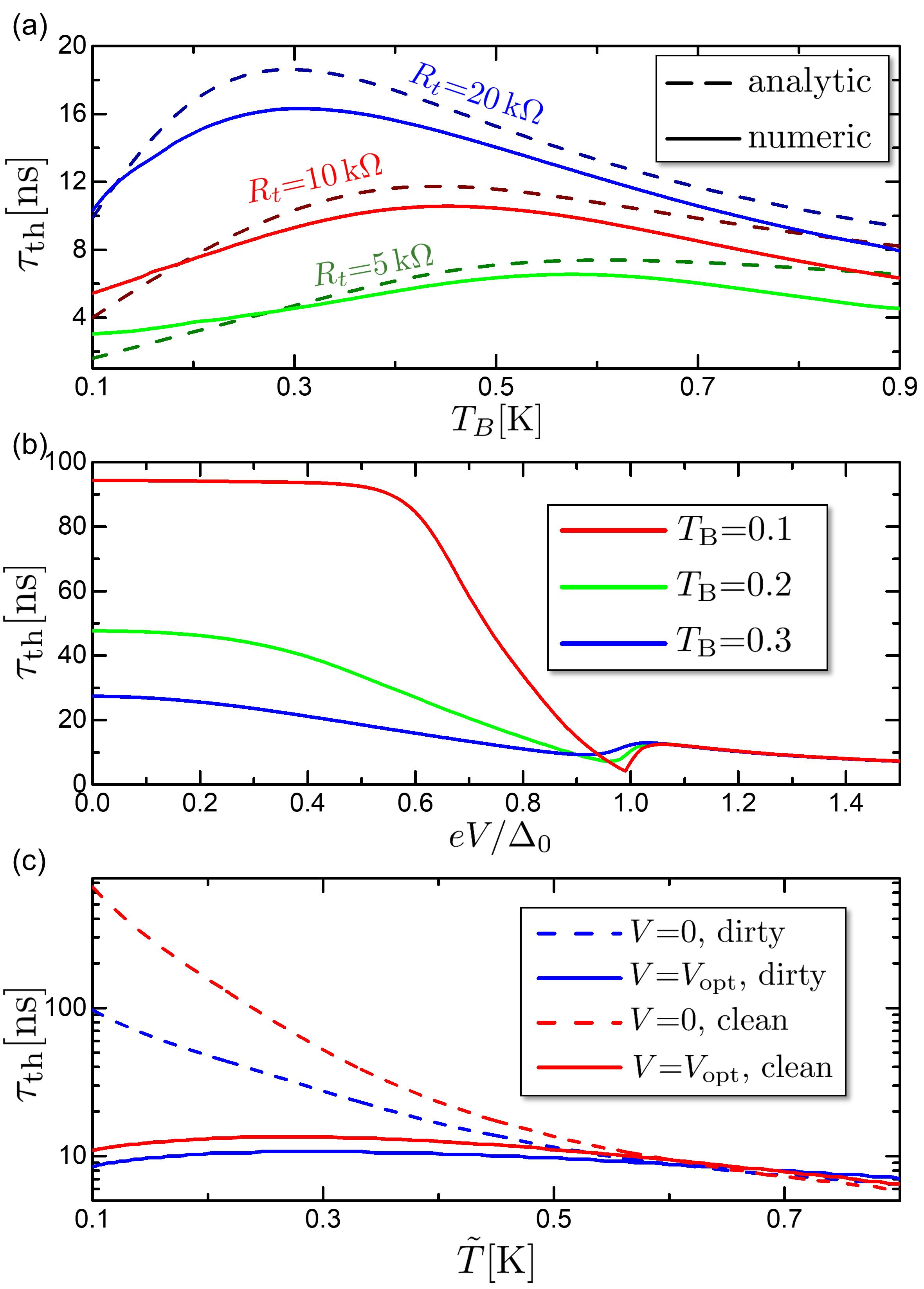}
	\caption{(a) Response time $\tauth$ versus $\TB$ at the optimal bias for a system with dirty graphene regime and different tunnel resistances $\Rt$, numerical and analytical (see Eq. (\ref{eq:TimeConstantAnalytics})).  (b) $\tauth$ vs $V$ for $\TB\text=$ $\rm0.1\,K, 0.3\,K, 0.5\,K$. (c) $\tauth$ vs $\tilde T$ in a system where $(V=0, \TG\text=\TB\text=\tilde T)$ (dashed curves) and in a system where $(V=\Vopt,\TGmin=\tilde T, \TB')$ where $\TB'$ is such that $\TGmin=\tilde T$ (solid curves).
	}
	\label{fig:tauVsParameters}
\end{figure}
Figure \ref{fig:tauVsParameters}a shows the dependence of $\tauth$ on $\TB$ at optimal voltage bias for different values of $\Rt$. The solid lines correspond to Eq. (\ref{eq:TimeConstantAnalytics}) where $\TGmin$ is given by Eq. (\ref{eq:AnalyticTG}). The dashed lines are obtained by numerically solving the heat balance equation when perturbing the graphene base temperature $\TG$ of 10\%. The numerical and analytical results are in good agreement. The maximum of each curve is the crossover point between the two regimes dominated by the junction [Eq. (\ref{eq:LowT})] and the e/ph coupling [Eq. (\ref{eq:HighT})]. The response time increases with $\Rt$, since the thermal conductance of the junction is lowered. In particular, at low $\TB$, the curves of Fig. \ref{fig:tauVsParameters}a indicate that $\tauth\propto \Rt$ as given by Eq. (\ref{eq:LowT}).

The results in Eq. (\ref{eq:TimeConstantAnalytics}) and Fig. \ref{fig:tauVsParameters}a  are obtained for $V=\Vopt$.  $\tauth$ has a dependence also on the bias voltage, since the latter tunes the transport properties of the junction. Figure \ref{fig:tauVsParameters}b reports $\tauth$ versus $V$ calculated for different bath temperatures in the case of dirty graphene  regime. We notice that the response time $\tauth$ decreases from $\rm95\,ns$ at $V=0$ to $\rm5\,ns$ at $V=\Vopt$ when $\TB=0.1\,$K, because when the cooling operates, the junction thermal conductance is enhanced. 

This point can be investigated analytically. To evaluate the voltage dependence of the thermal response at small bias, we need the thermal conductance of the junction $\GGIS (\TG=\TB,V=0)=\partial_{\TG}\ThC(\TG=\TB,V=0)$. It can be approximated by the tunnel integral expression in Eq. (\ref{eq:GGISBiased}) at $\kB \TG, \kB \TB\ll \Delta_0$. At the leading order we obtain finally
\begin{equation}
G_{\rm GIS}(\TG,V=0) \sim  \frac{\sqrt{2 \pi}\Delta_0 \kB}{e^2\Rt}\left(\frac{\kB \TG}{\Delta_0}\right)^{-3/2}e^{-\Delta_0/\kB \TG}  \, \, .
\label{eq:GGISZeroBias}
\end{equation}
Linearizing the heat balance equation around the equilibrium state $(\TG=\TB,V=0)$ we obtain 
\begin{equation}
\tauth = \frac{C}{G_{\rm tot}} \approx \frac{A \gamma \TGmin}{\delta A \Sigma_\delta \TGmin^{\delta-1} + \GGIS (\TG\text=\TB,V\text=0) } \, \, .
\label{eq:TimeConstantAnalyticsNoBias}
\end{equation}
The difference between $\tauth(V=0)$ [Eq. (\ref{eq:GGISZeroBias})] and $\tauth(V=\Vopt)$ [Eq. (\ref{eq:GGISBiased})] is strong. In particular, at low temperatures the junction conductance is exponentially suppressed at zero bias, while $\GGIS$ has a large contribution in the optimally biased case.

The difference of $\tauth$ between the biased and unbiased case is remarked in Fig. \ref{fig:tauVsParameters}c. Dashed curves show $\tauth$  in an unbiased system at $(\TG\text=\TB\text=\tilde T, V\text=0)$, while solid curves show $\tauth$ for $\TGmin = \tilde T$ and $\TB, \Vopt(\TB)$ are set subsequently. For completeness, we show both the dirty (blue curves) and clean (red curves) graphene  regimes. The difference in response time between $V=0$ and $V=\Vopt$ can reach one or two orders of magnitude depending on the value of $\TG$ and the graphene regime. Furthermore, at $V=0$, there is no maximum in $\tauth$, since both the $\Geph$ and $\GGIS$ are increasing functions of $\TG$.

It is worth to note that the response time does not depend on carrier density $n$. Indeed, both $C$ and $G_{\rm tot}$ are proportional to $\sqrt{n}$. As a consequence, the gating does not affect $\tauth$.

Finally, we evaluate the temperature response to a finite external power signal $\ThCin\neq 0$. This quantity will be exploited for investigating the bolometric response of the device. It is useful to write the linear heat balance equation (\ref{eq:LinearBalanceEquation}) in the frequency domain including the signal $\ThCin(\omega)$. We remark that the frequency $\omega$ of $\ThCin$ refers to the Fourier component of the power and not to the electromagnetic frequency. The resulting equation takes the form
\begin{equation}
\Delta \TG (\omega) = {\cal T}_{TP}(\omega)\ThCin(\omega) = \frac{1}{G_{\rm tot}(1+i\omega \tauth)} \ThCin(\omega) \, \, ,
\label{eq:Power2Temperature}
\end{equation}
where ${\cal T}_{TP}=1/(G_{\rm tot}(1+i\omega \tauth))$ is the power-to-temperature transfer function. This equation shows that the SIGIS responds as a low-pass filter with cut-off frequency $\omega_0=1/\tauth$. Considering the values of $\tauth$ reported in Fig. \ref{fig:tauVsParameters}a, the corresponding frequency is in a range {of $\SI{10}{\mega\hertz}-\SI{60}{\mega\hertz}$}.
In the following section, this transfer function will be used to evaluate the responsivity, a figure of merit which quantifies the SIGIS performances as a bolometer. 

\section{Biased SIGIS as a bolometer}
\label{sec:Bolometer}
In this section, we study the cooled SIGIS as a bolometer. An input power $\ThCin$ is converted in a variation of current when the SIGIS is kept at a constant voltage bias. In detail, we characterize two bolometric figures of merit, the responsivity and the NEP.

The bolometric properties of a SINIS system with electron cooling have been studied in literature \cite{golubev2001,kuzmin1998,lemzyakov2018,kuzmin2019}. The main result is that the built-in refrigeration enhances the responsivity and decreases the NEP. Here, we essentially follow a similar analysis for a SIGIS.

We point out that SIGIS systems have already been  investigated in literature, at $V\to0$, where the cooling is negligible \cite{vora2012,mckitterick2015,du2014}. The purpose of these low $V$ schemes is to decrease the thermal conductance across the junction in order to use the device at lower input power regimes \cite{vora2012,mckitterick2015,du2014}.

\begin{figure}[t]
\centering
\includegraphics[width=0.48\textwidth]{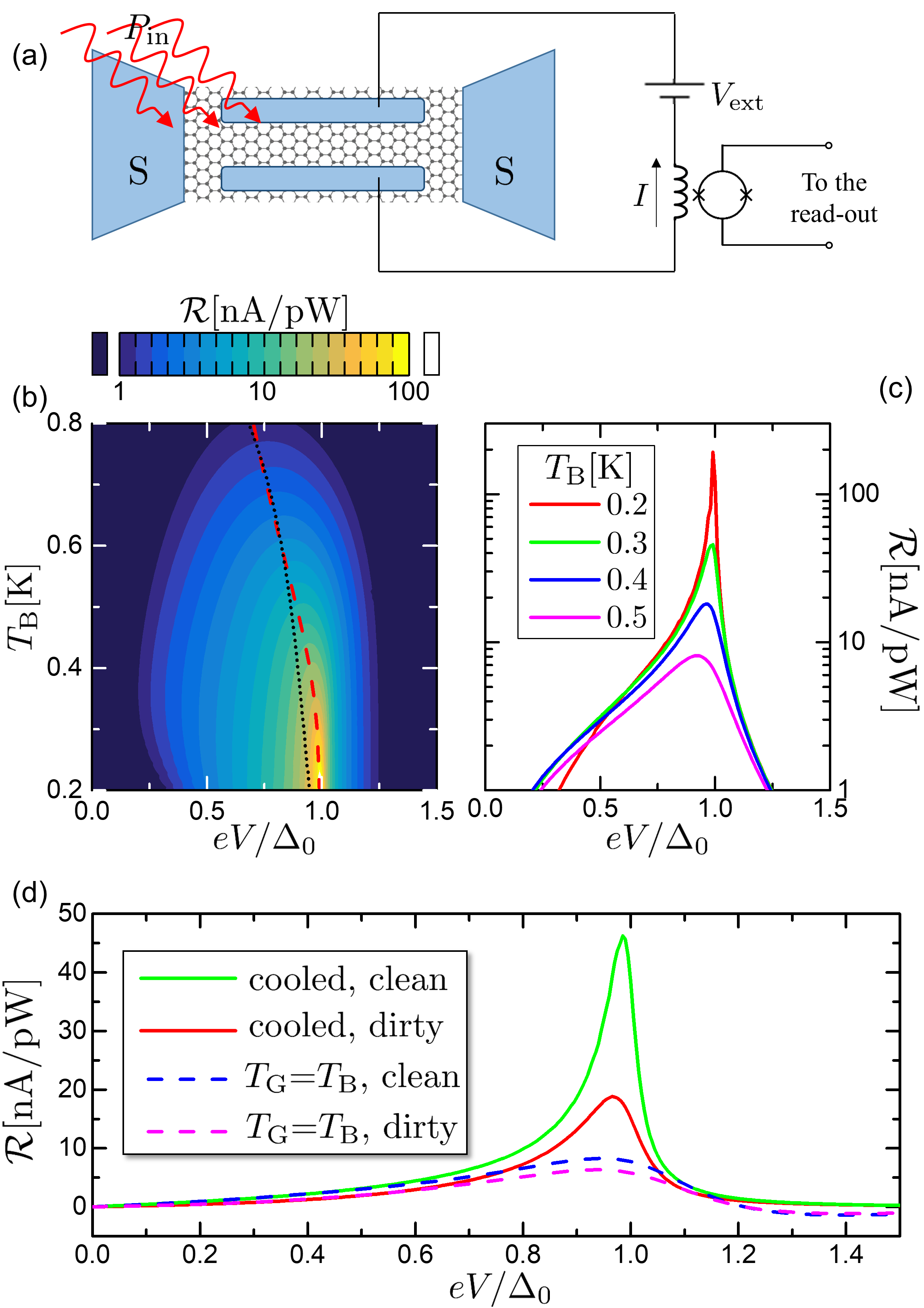}
\caption{
(a) Bolometric detection scheme for the SIGIS system. The graphene is in clean contact with a superconducting antenna. A photonic power $\ThCin$ increases the temperature of the electrons in graphene and changes the tunneling rate across the GIS junctions, resulting in a variation of the current $I$. The current is detected and amplified by a superconducting interferometer. (b) Color map of the responsivity $\Resp$ versus $(V,\TB)$. (c) Cuts from panel (b) for the chosen temperatures in legend. (d) $\Resp$ in  the cases of dirty and clean graphene  regime, in the presence of cooling or for $\TG=\TB$ (inefficient cooling), see legend. The curves are obtained at $\TB=0.3\,$K.
}
\label{fig:Responsivity}
\end{figure}
Our bolometer scheme consists of a SIGIS system connected to an external voltage generator $V_{\rm ext}=2V$, being $V$ the voltage drop across a single junction (see sketch in Fig. \ref{fig:Responsivity}a). The graphene is also connected to the superconducting antenna by means of a clean superconductor/graphene junction. The superconducting antenna allows carrying the power $\ThCin$ and traps it in graphene since the superconducting leads work as Andreev mirrors  \cite{golubev2001,nahum1993ApplPhysLett}, reducing the thermal leakage to the antenna. It is important to remark that the distance between the antenna electrodes must be enough to make the Josephson coupling through proximity effect  negligible \cite{heersche2007}. The electric current $I$ in the circuit is measured by means of an inductance coupled to a superconducting interferometer read-out  \cite{clarke2005,giazotto2010, giazotto2011,ronzani2014,dambrosio2015}.  

\subsection{Responsivity}
We start our investigation with the responsivity, defined as a power-to-current transfer function:
\begin{equation}
\Resp(\omega) = \frac{\partial I(\omega)}{\partial \ThCin (\omega)} \, \, ,
\end{equation}
where $I(\omega)$ and $\ThCin(\omega)$ are the electric current and the input power signal in the frequency domain, respectively.

We calculate the responsivity as the product of the power-to-temperature transfer function ${\cal T}_{TP}$ in Eq. (\ref{eq:Power2Temperature}) with the temperature-to-current transfer function ${\cal T}_{IT}= \partial_{\Delta T_{\rm G}}I$. The product of the two transfer functions is equivalent to calculate the derivative $\Resp = \partial_{{\ThCin}}  I$ by the factorization $\Resp = \partial_{{\TG}} I \times (\partial_{{\TG}}\ThCin)^{-1} $, since ${\cal T}_{TP} = \partial_{\ThCin} \Delta T$ \cite{golubev2001}. We obtain 
 \begin{equation}
\Resp(\omega) = {\cal T}_{IT}(\omega) {\cal T}_{TP}(\omega) = \frac{\partial I/\partial \TG }{G_{\rm tot}(1+i\omega \tauth)} \ThCin(\omega) \, \, .
\label{eq:GeneralResponsivity}
\end{equation}
The responsivity has a cut-off at the frequency $\omega_0=1/\tauth$.

We focus on the low frequency limit, which is valid when the band of the input signal is sufficiently below the cut-off frequency. Fig. \ref{fig:Responsivity}b reports a color map of $\Resp$ versus $V$ and $\TB$, obtained by Eq. (\ref{eq:GeneralResponsivity}) using the numerical derivative of Eqs. (\ref{eq:ElectricalCurrent}), (\ref{eq:ThermalCurrent}). Cuts of Fig. \ref{fig:Responsivity}b versus $V$ are reported in Fig. \ref{fig:Responsivity}c. The responsivity shows a peak on the red dashed curve $V_{\rm opt}^{\Resp}(\TB)$. The latter does not coincide with $\Vopt$ (dotted black in Fig. \ref{fig:Responsivity}b), which maximizes the cooling performances. Indeed,  $V_{\rm opt}^{\Resp}(\TB)$ and $\Vopt$ are different by definition, since the former is obtained by maximizing $\partial_{T_{\rm G}} I/\partial_{ T_{\rm G}} \ThCin$ and the latter by maximizing $\ThC$. $V_{\rm opt}^{\Resp}(\TB)$ is located closely below $\Delta(\TB)/e$. Above this voltage, the current characteristics $I(V,\TG,\TB)$ lose sensitivity to temperature since they converge to the ohmic behavior $I=V/\Rt$. On the other hand, for $V$ well below the gap, the current is suppressed. 

Other physical features of responsivity are represented in Fig. \ref{fig:Responsivity}d. Here,  the solid curves are calculated by considering  the graphene cooling, while the dashed curves are obtained by imposing $\TGmin=\TB$, i.e., disregarding the cooling effect. This treatment corresponds to a physical situation where a spurious heating source completely spoils the cooling power of the junction. Let us investigate how the difference of graphene  regime affects the responsivity. We first consider the dashed curves in Fig. \ref{fig:Responsivity}d, representing the absence of cooling, where we can notice that the clean case is slightly more responsive. The reason is due to the enhanced power-to-temperature transfer function ${\cal T}_{TP}$. Indeed, in both the dashed results ($\TGmin=\TB$), the temperature to current transfer function ${\cal T}_{IT}$ in Eq. (\ref{eq:GeneralResponsivity}) is the same, since it is a property of the junction depending only on $V,\TG$, and $\TB$. But the transfer function ${\cal T}_{TP}$ changes between a clean or dirty graphene regime, since the phonon thermal conductance is lower in the clean case. This means that, given a power input, the temperature raise $\Delta \TG$ is bigger in the clean case, resulting in a greater current response.

The comparison between the dashed and solid curves in Fig. \ref{fig:Responsivity}d shows that the presence of an active cooling enhances the responsivity. The graphene base temperature is lower for clean graphene regime (see Sec. \ref{sec:TGmin}), resulting in a stronger enhancement of responsivity compared to the dirty graphene case.

A physical insight to this argument can be obtained by using the low temperature approximations studied above. We underline that these expressions hold for $\Vopt$ and not $\Vopt^{\cal R}$, but they give enough information for a physical picture. The responsivity at low temperatures is 
\begin{equation}
\Resp = \frac{0.24\frac{\kB}{e \Rt} \left( {\frac{\kB \TGmin}{\Delta_0}}\right)^{-1/2} }{\delta A \Sigma_\delta \TGmin^{\delta-1} + 1.8 \frac{\Delta_0 \kB}{e^2 \Rt} \left(
\frac{\kB \TGmin}{\Delta_0} 
\right)^{1/2} } \, \, .
\label{eq:ResponsivityAnalytic}
\end{equation}
As in the previous section, the denominator shows the presence of two regimes separated by the crossover temperature $\Tcross$ in Eq. (\ref{eq:TCrossover}). The regime at $\TGmin \gg \Tcross$ is dominated by the e/ph thermal channel with responsivity
\begin{equation}
\Resp\approx \frac{0.24 \sqrt{\kB \Delta_0}}{e \Rt \delta A \Sigma_\delta \TGmin^{\delta-0.5}}   \, \, .
\end{equation}
The regime at $\TGmin \ll \Tcross$ is dominated by the junction thermal channel with responsivity at $\Vopt$
\begin{equation}
\Resp \approx 0.13 \frac{e}{\kB \TGmin} \approx \frac{0.15}{\TGmin{\rm [K]}}{\rm \frac{nA}{pW}} \,\, .
\label{eq:RespBelowCrossover}
\end{equation}
This expression does not involve any graphene property, but it is obtained by the ratio of the two junction properties $\partial_{T_{\rm G}} I$ and $\GGIS = \partial_{T_{\rm G}} \ThC$. In particular, both terms scale as $1/\Rt$, so the tunnel resistance does not directly affect the responsivity at low temperatures.

Finally, we would like to stress that the responsivity increases by decreasing the graphene temperature. This is also confirmed  by Fig. \ref{fig:Responsivity}b,c.

\subsection{Noise equivalent power}
\begin{figure}[t]
	\centering
	\includegraphics[width=0.48\textwidth]{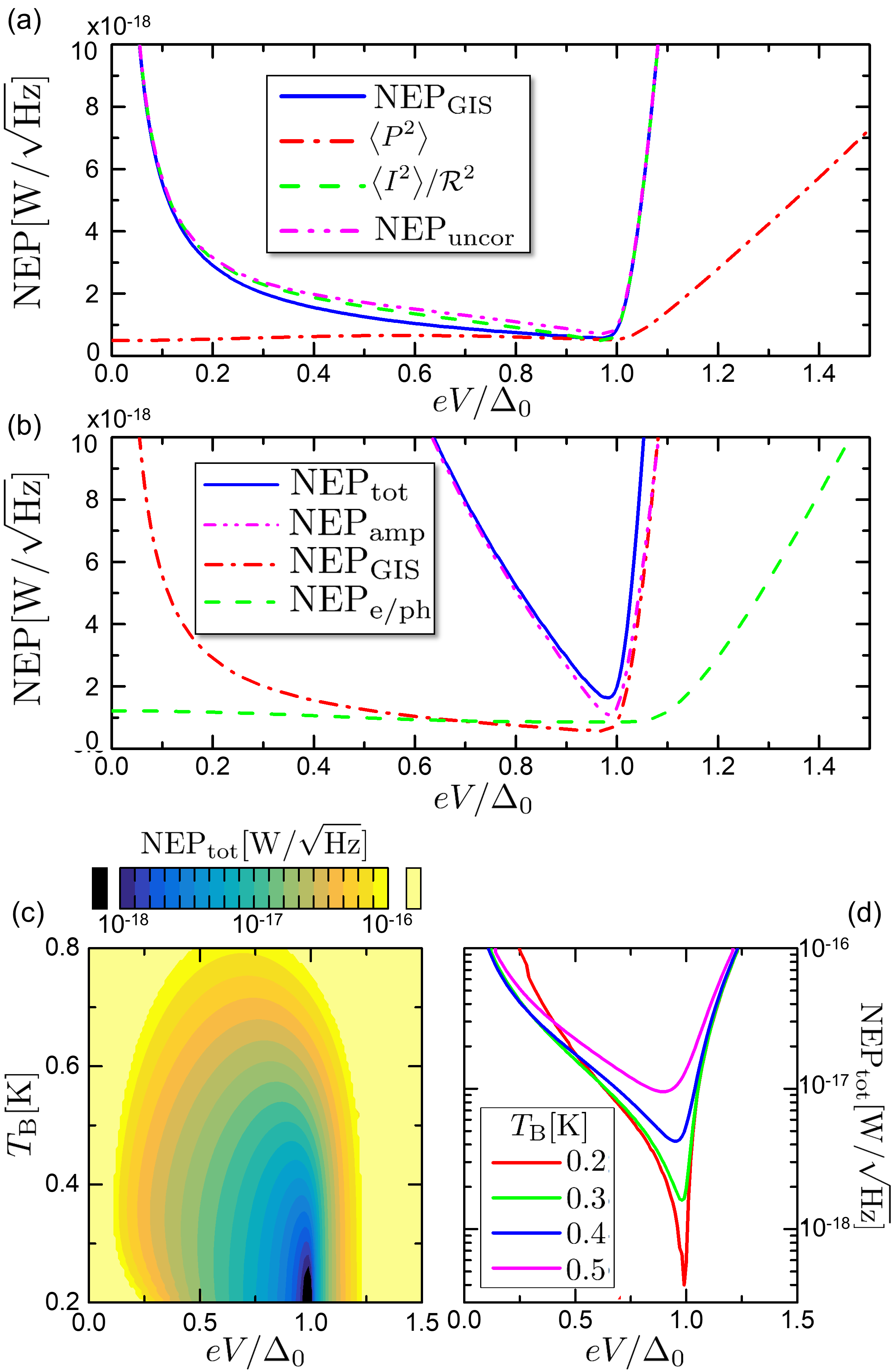}
	\caption{(a) NEP components in a GIS junction at $\TB=0.3\,$K (clean graphene regime). (b) NEP components to the total NEP (clean graphene regime). (c) NEP color map versus $V$ and $\TB$. (d) NEP cuts from panel (c) at bath temperatures $\TB$ in legend.
	}
	\label{fig:NEPs}
\end{figure}
We now focus on the noise equivalent 
power, which is defined as the signal power necessary to have a signal-to-noise ratio equal to 1 with a bandwidth of 1 Hz \cite{richards1994}. 

The total NEP of the SIGIS is given by different contributions \cite{golubev2001}
\begin{equation}
\NEP_{\rm tot}^2 = 2 \NEP_{\rm GIS}^2  +  \NEP_{\rm e/ph}^2 +\NEP_{\rm amp}^2 \, \, ,
\label{eq:TotalNEP} 
\end{equation}
where the three terms are related to the junction, the e/ph coupling and the amplifier read-out, respectively. The factor 2 in front of $\NEP_{\rm GIS}^2$ takes into account the two junctions, assuming their noises to be uncorrelated \cite{giazotto2005}, which is related to the fact that temperature fluctuations, as the one induced by heat noise, are small in comparison to the stationary value of $\TG$ \cite{guarcello2019}.

The contribution to the junction NEP is given by fluctuations in both the electric and heat currents:
\begin{equation}
\NEP_{\rm GIS}^2 = \PNoise-2\frac{\IPNoise}{\Resp}+\frac{\INoise}{\Resp^2} \, \, ,
\label{eq:JunctionNEP}
\end{equation}
where the quantities in angled brackets are the low frequency spectral densities of fluctuations \cite{golubev2001}. $\INoise$ is the current fluctuation given by 
\cite{golubev2001} 
\begin{multline}
\INoise = \frac{2}{\Rt} \int_{-\infty}^{\infty} \dd \en\left\lbrace  \DoSG(\en\text-eV\text-\uG) \DoSS(\en) \times \right.\\
\left.  [f(\en\text-eV,\TG)+f(\en,\TS)-2f(\en\text-eV,\TG)f(\en,\TS)] \right\rbrace \, \, .
\label{eq:INoise}
\end{multline}
The fluctuation of the tunneling rate is mirrored in a fluctuation $\PNoise$ of the tunneled heat
\begin{multline}
\PNoise = \frac{2}{e^2 \Rt} \int_{-\infty}^{\infty} \dd \en \left\lbrace 
(\en\text-eV)^2 \DoSG(\en\text-eV\text-\uG) \DoSS(\en) \times \right.\\
\left.  [f(\en\text-eV,\TG)+f(\en,\TS)-2f(\en\text-eV,\TG)f(\en,\TS)] \right\rbrace \, \, .
\label{eq:PNoise}
\end{multline}
Since the two fluctuations $\INoise$ and $\PNoise$ are given by the tunneling of the same carriers, a non-null correlation exists \cite{golubev2001}: 
\begin{multline}
\IPNoise = \frac{2}{e \Rt} \int_{-\infty}^{\infty} \dd \en\left\lbrace 
(\en\text-eV) \DoSG(\en\text-eV\text-\uG) \DoSS(\en) \times \right.\\
\left.  [f(\en\text-eV,\TG)+f(\en,\TS)-2f(\en\text-eV,\TG)f(\en,\TS)] \right\rbrace \, \, .
\label{eq:IPNoise}
\end{multline}
In these integrals, the energy dependence of graphene has been neglected, according to the approximation done in Sec. \ref{sec:model}.

Figure \ref{fig:NEPs} reports the NEP components for $\TS=\TB=0.3\,$K. Panel (a) shows the contributions to $\NEP_{\rm GIS}$ in Eq. (\ref{eq:JunctionNEP}). For completeness, the NEP calculated by neglecting the cross-correlation between $\INoise$ and $\PNoise$ is also reported 
\begin{equation}
\NEP_{\rm unc} =  \PNoise + \frac{\INoise}{\Resp^2} \, \, .
\end{equation}
By comparing $\NEP_{\rm unc}$ and $\NEP_{\rm GIS}$ we can notice that the $\IPNoise$ term brings a correction that reduces the total NEP. The cross-correlation is positive except in the region above the gap voltage $\Delta/e+0.6\kB \TG/e<V<\Delta/e+1.3 \kB \TG/e$. Outside this region, the cross-correlation partially cancels the shot noise and the heat noise \cite{golubev2001}.

The NEP due to the junction noise is smaller in a SIGIS bolometer compared to a SINIS bolometer. Indeed, $\NEP_{\rm GIS}$ scales as $\Rt^{-1/2}$ and good cooling characteristics can be reached in a SIGIS with a tunnel resistance one order of magnitude greater compared to a SINIS. As a consequence, the $\NEP_{\rm GIS}$ is lower of a factor $\sim 3$.

Let us consider the other NEP contributions. The contribution related to the noise in the e/ph channel can be roughly estimated by a generalization of expression in Ref. \cite{golubev2001}
\begin{equation}
\NEP_{\rm e/ph}^2 = 2\delta \kB A\Sigma_\delta (\TGmin^{\delta+1}+\TB^{\delta+1})\, \, .
\label{eq:PhononNEP}
\end{equation}
At equilibrium $\TG=\TB=T$, the NEP takes the standard form $\NEP_{\rm e/ph}^2=4\kB \Geph T^2$ \cite{mckitterick2015,du2014}. We notice that this term is smaller in a SIGIS compared to a SINIS, due to the lower e/ph coupling constant (see discussion in Sec. \ref{sec:TGmin}). In the temperature range of 0.1K-1K, the e/ph thermal conductance is one order of magnitude lower, yielding a $\NEP_{\rm e/ph}$ decrease of a factor $\sim 3$.

Finally, we consider the read-out NEP due to the amplifier noise $\INoise_{\rm amp}$
\begin{equation}
\NEP_{\rm amp}^2 = \frac{\INoise_{\rm amp}}{\Resp^2}
\label{eq:AmplifierNEP}
\end{equation}
and we assume $\sqrt{\INoise_{\rm amp}}\approx{\rm 0.05 \, pA/\sqrt{Hz}}$ \cite{golubev2001}.

Panel (b) of Fig. \ref{fig:NEPs} shows the different contributions to the total NEP at $\TB=0.3\,$K versus $V$. Panels (a) and (b) show the same $\NEP_{\rm GIS}$. We notice that $\NEP_{\rm tot}$  has a minimum close to the optimal bias. Here, the three contributions are of the same order of magnitude and yield $\NEP_{\rm tot} = \SI{1.6E-18}{\watt/\sqrt{\hertz}}$. Away from the optimal point, the read-out $\NEP_{\rm amp}$ dominates. Hence, in order to optimize the total NEP, it is important to reduce the noise of the measurement circuitry.

The electronic cooling influences the NEP in two ways: on one side, it decreases the  thermal fluctuations of electrons in graphene, on the other it enhances the responsivity (see Fig. \ref{fig:NEPs}b). The former effect is quantified by the low temperature expressions  $\Vopt$ $\INoise\approx (\kB \TGmin)^{1/2}\sqrt{\Delta_0}/e \Rt$, $\IPNoise\approx (\kB \TGmin)^{3/2}\sqrt{\Delta_0}/e \Rt$, $\PNoise\approx (\kB \TGmin)^{5/2}\sqrt{\Delta_0}/e^2 \Rt$ \cite{golubev2001}. The latter effect involves all the contributions that have $\Resp$ at the denominator. This is remarked by the total NEP versus $(V,\TB)$ shown in Fig. \ref{fig:NEPs}c,d, that resembles the inverse of responsivity in panels \ref{fig:Responsivity}b,c. {In particular, the NEP improves of about two orders of magnitude moving from the zero-bias to the optimal-bias configuration.}

We now investigate the effects of the carrier density $n$ on the bolometric properties. The responsivity is not affected by $n$, since ${\cal T}_{TP}\propto G_{\rm tot}^{-1}\propto n^{-1/2}$ and ${\cal T}_{IT}\propto \Rt^{-1}\propto n^{1/2}$. The term $\NEP_{\rm GIS}\propto\Rt^{-1/2}\propto n^{1/4}$ and similarly $\NEP_{\rm e/ph}\propto \Sigma_{\delta}^{1/2}\propto n^{1/4}$. The read-out term instead does not depend on $n$. Hence, the NEP is a weakly increasing function of $n$. Considering that the gating can vary $n$ from the residual charge $n_0$ of a factor $100$ at most, the NEP can vary of a factor $\sim 3$. Therefore, the bolometric properties can be considered stable under charge variations or fluctuations.

\section{Dependence on Dynes parameter}
\label{sec:GammaDynes}
\begin{figure}[t]
	\centering
	\includegraphics[width=0.48\textwidth]{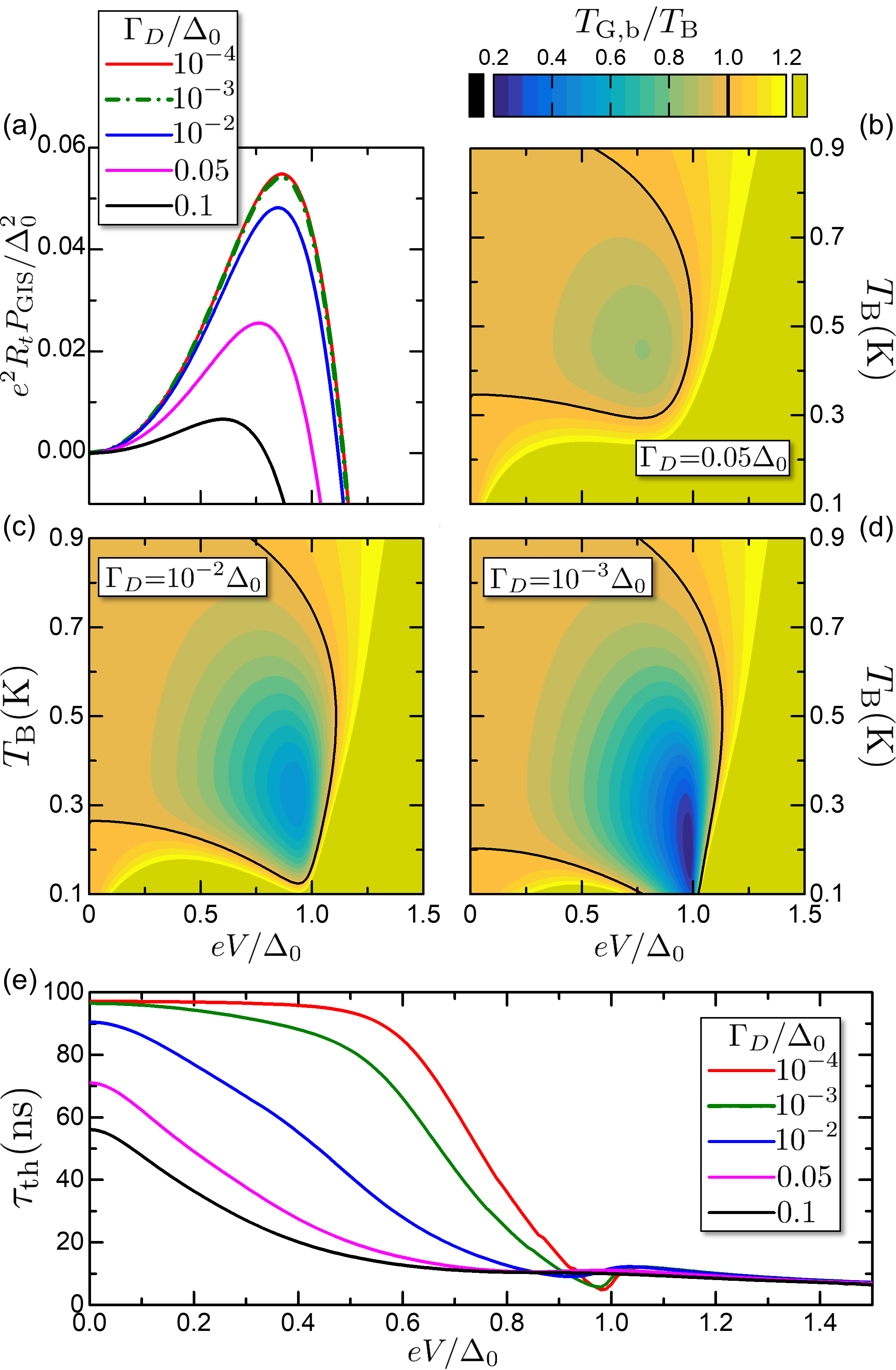}
	\caption{(a) Cooling power at $\TG=\TB=0.5\,$K versus the bias $V$, for different values of $\Gamma_D$ in legend. (b-d) Contour plots of the ratio $\TGmin/\TB$ versus bias $V$ and bath temperature $\TB$, for the values $\Gamma_D/\Delta_0=0.05,10^{-2},10^{-3}$. (e) Response time $\tauth$ versus $V$ at $\TB=0.1\,$K for the values of $\Gamma_D$ in legend, dirty e/ph coupling.
	}
	\label{fig:DynesOnTGb}
\end{figure}
Let us discuss here the role of the Dynes parameter, introduced in Eq. (\ref{eq:SuperconductorDoS}). This phenomenological parameter takes into account the finiteness of the superconducting peaks and the subgap tunneling \cite{dynes1978}. The latter strongly depends on different issues, e.g., the fabrication quality of the junction \cite{herman2016} and, more generally, on environmental effects \cite{pekola2010}. For this reason, $\Gamma_D$ is frequently used as a parameter to quantify the quality of a tunnel junction with a superconductor. Indeed, realization of high-quality tunnel junctions is an important requirement to avoid effective sub-gap conduction channels. The value of $\Gamma_D$ can be extracted experimentally from a fit of the measured electrical differential conductance $G_e(V)$ at low temperature $\kB\TG,\kB\TB\ll \Delta_0$, where $G_e(V)\propto \rho_{\rm S}(eV)$.  

The sub-gap density of states is
\begin{equation}
\rho_{\rm S} (\en < \Delta_0)\simeq \frac{\Gamma_D}{\Delta_0} \, \, ,
\end{equation}
which implies that for $eV,\kB\TG,\kB\TB \ll \Delta_0$ the junction behaves as a NIN with effective resistance $\tilde{\Rt} =\Rt \Delta_0/\Gamma_D$, with current $I\simeq V/\tilde{\Rt}$ and Joule heating $V^2/\tilde{\Rt}$.

In the previous sections, we assumed good quality junctions with $\Gamma_D=10^{-4}\Delta_0$. Such a value of $\Gamma_D$ has been experimentally realized in metallic NIS junction, while it has not been reached in graphene junctions yet. Quality of GIS junctions is improved over time, and it can be nowadays expressed by $\Gamma_D$ on the order of $10^{-1}\Delta_0$. State of the art experiments hint that $\Gamma_D\approx\SI{7E-2}{}\Delta_0$ \cite{zihlmann2018}. 

In this section, we show how the Dynes parameter affects cooling and bolometric characteristics.

{\it Effects on cooling.} The cooling power is reduced by the increasing of the Dynes parameter since the smearing of the peaks in the BCS-Dynes DoS does not allow sharp filtering of the hot electrons \cite{giazotto2006,muhonen2012,pekola2004}. Moreover, the sub-gap conduction implies a Joule heating $V^2/\tilde{\Rt}$, half of which flows in graphene. Figure \ref{fig:DynesOnTGb}a shows the cooling power $\ThC$ versus the bias for different values of $\Gamma_D$, at the temperature $\TG=\TB=0.5\,$K. Up to $\Gamma_D=10^{-2}\Delta_0$, the cooling power is slightly affected by $\Gamma_D$. From $\Gamma_D=10^{-2}\Delta_0$ to $\Gamma_D=10^{-1}\Delta_0$, the cooling power is strongly decreased. This is mirrored in the graphene base temperature $\TGmin$. Panels (b,c,d) of Fig. \ref{fig:DynesOnTGb} show $\TGmin/\TB$ versus the bias $V$ and the bath temperature $\TB$, for $\Gamma_D/\Delta_0 = 0.05, 10^{-2},10^{-3}$, respectively. In particular, the region of $(V,\TB)$ where the temperature is decreased depends on $\Gamma_D$. Anyways, the simulations suggest that cooling can still be observed  for $\Gamma_D=0.05\Delta_0$, where $\TGmin/\TB$ can reach the value of $\sim0.8$. For $\Gamma_D=10^{-2}\Delta_0$, the cooling is well operating.  For $\Gamma_D=10^{-3}\Delta_0$, the $\TGmin/\TB$ plot resembles the one in Fig.  \ref{fig:CoolingSimulation}a.

{\it Effects on the response time.} The value of $\tauth$ is weakly affected by $\Gamma_D$ at $\Vopt$. Indeed, when the junction is biased, the sub-gap contribution to the thermal conductance plays a marginal role compared to the contribution of the states above the gap. In Fig. \ref{fig:DynesOnTGb}e, we report instead what happens at finite bias, plotting $\tauth$ at $\TB=0.1\,\rm K$ versus $V$ for different values of $\Gamma_D$.  At $e V\sim \Delta_0$, the response time is weakly affected by $\Gamma_D$, keeping on the order of 10 ns. The response time is affected by $\Gamma_D$ only around $V\sim0$ and at low temperatures $\TG,\TB \lesssim 0.2\,\rm K$, since the contribution of the sub-gap conduction and the electron-phonon coupling are comparable. Anyways, we remark that for $\TG,\TB \gtrsim 0.2\,\rm K$, the dependence on $\Gamma_D$ is negligible, independently on the bias $V$.

\begin{figure}[t]
	\centering
	\includegraphics[width=0.48\textwidth]{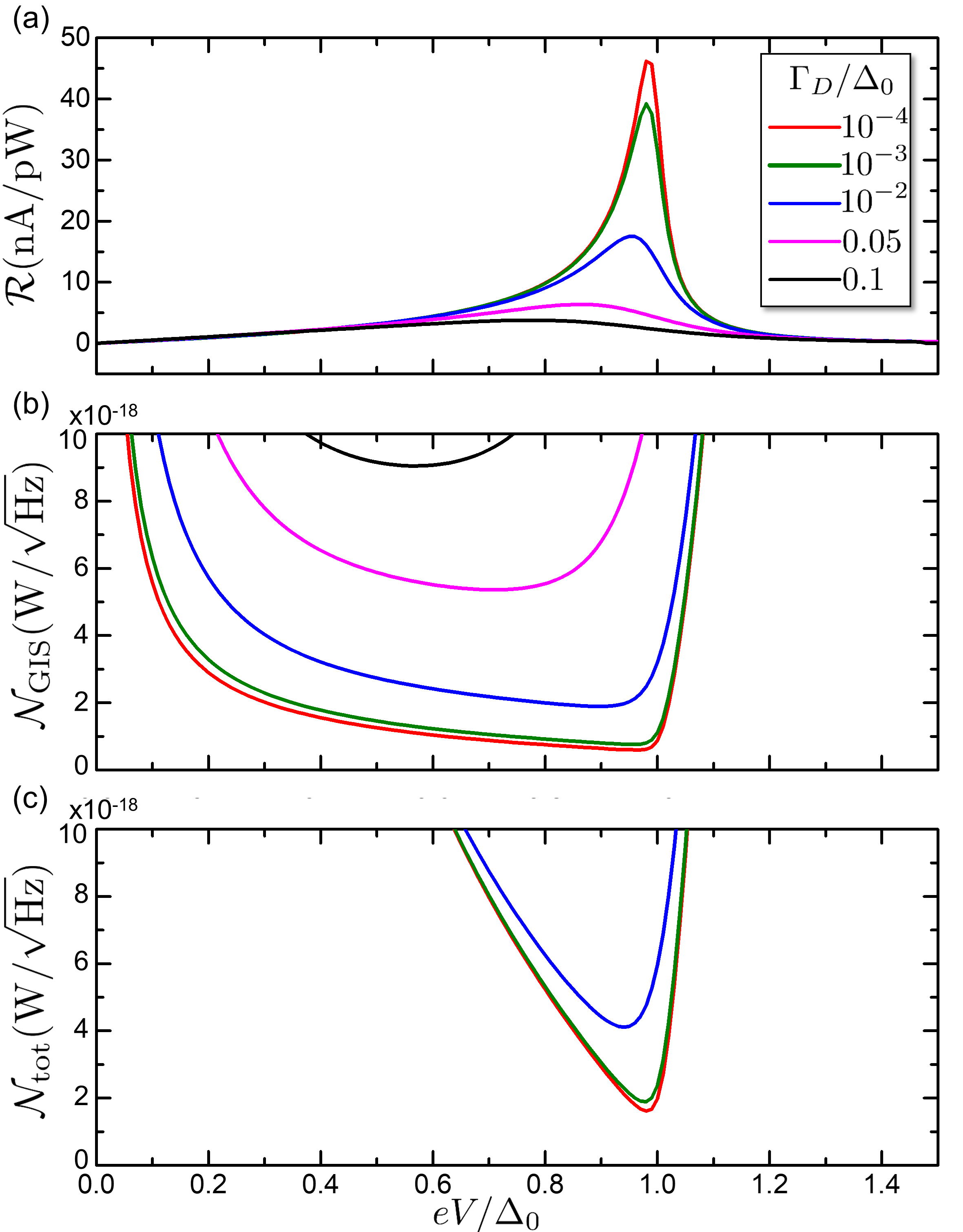}
	\caption{Bolometric characteristics versus bias $V$ for different values of the Dynes parameter $\Gamma_D$, in legend. (a) Responsivity $\Resp$ at $\TB=0.3\,$K, clean e/ph coupling. (b) Junction NEP at $\TB=0.3\,$K. (c) Total NEP at $\TB=0.3\,$K, clean e/ph coupling. The NEP at $\Gamma_D/\Delta_0 = 5\times10^{-2}, 10^{-1}$ is out of scale, with respective minimum values $\simeq \rm 1.1 \times 10^{-17} W/\sqrt{Hz}$ and $\simeq \rm 1.9 \times 10^{-17} W/\sqrt{Hz}$.
	}
	\label{fig:BolomVsGamma}
\end{figure}
{\it Effects on responsivity.} The value of $\Resp$ is affected by $\Gamma_D$ through the $\TGmin$ increase and, at the same time, by the reduction of $\partial I /\partial T$, since the smeared DoS peaks are translated in less sharp features of the $I(V)$ characteristics in temperature. Figure \ref{fig:BolomVsGamma}a shows $\Resp$ versus $V$ for different values of $\Gamma_D$, for $\TB=0.3\,$K and clean e/ph coupling. The peak of $\Resp$ decreases by a factor 0.38 at $\Gamma_D=10^{-2}$ and about one order of magnitude at $\Gamma_D=0.05\Delta_0$. 

{\it Effects on NEP.} The behavior of $\Resp$ on $\Gamma_D$ is reflected in the NEP characteristics. Indeed, $\Resp$ is present in the denominators of the NEP components in Eqs. (\ref{eq:JunctionNEP}) and (\ref{eq:AmplifierNEP}), while the numerators are weakly affected by $\Gamma_D$ at $eV\simeq\Delta_0$. Panels (c) and (d) of Fig. \ref{fig:BolomVsGamma} report the single junction $\NEP_{\rm GIS}$ and the total NEP at $\TB=0.3\,$K, calculated in the same manner of section \ref{sec:Bolometer}. Like the responsivity, the NEP worsen one order of magnitude to $\Gamma_D=0.05\Delta_0$. 

In summary, in this section, we have shown that the quality of the GIS junctions might play a role in the characteristics of the studied device. In particular, the Dynes parameter is detrimental for cooling and bolometric applications only when $\Gamma_D \gtrsim 10^{-2}\Delta_0$. 

\section{Comparison with other bolometric architectures}
\label{sec:comparison}
Bolometric technology is a very wide topic, stimulated mainly by challenges in astroparticle physics, e.g., study of the cosmic microwave background \cite{crittenden1993,seljak1997} or axion detection for dark matter investigation \cite{krauss1985,graham2015,hochberg2016,cast2017}.
The differences among bolometers concern many experimental features, such as fabrication issues, working temperature, read-out schemes, figures of merit. Among all the different characteristics, detectors combining low noise with fast response speed are highly desirable. Nevertheless, in bolometers technology, there is a trade-off between NEP and response time. Indeed, a fast response time is associated with a fast heat dissipation through thermal channels. However, a large thermal dissipation corresponds to a low responsivity and to a large thermal coupling with external systems, both deteriorating the NEP. Hence, in an experimental setup, it is important to choose the right compromise between $\tauth$ and the NEP $\NEP$ on the base of the specific requirements.

A comparison based on the various experimental features of all the different bolometric technologies  is beyond the scope of this article. Here, we compare our SIGIS with three bolometric architectures, similar in working principles or materials. The first architecture concerns SINIS bolometers with built-in electron refrigeration  \cite{nahum1993ApplPhysLett,nahum1993IEEE,nahum1995,kuzmin1998,golubev2001,lemzyakov2018,kuzmin2019}. Second, we consider SIGIS bolometers based on power-to-resistance conversion at $V=0$ bias \cite{vora2012,mckitterick2015,du2014}. Finally, we consider also bolometers based on proximity effect in SNS  \cite{govenius2014,govenius2016,kokkoniemi2019} and SGS junctions \cite{walsh2017,lee2019}. 

{\it SINIS bolometers.} Similarly to our device, SINIS bolometers exploit the capability of a voltage bias to provide both cooling and extraction of the bolometric current signal. The theoretical work in Ref. \cite{golubev2001} predicts $\tau \sim \SI{0.2}{\micro\second}$ and  $\NEP\sim\SI{4E-18}{\watt/\sqrt{\hertz}}$ at temperature $\sim300\,$mK. Recent experiments have shown a response time $\tau \sim \SI{2}{\micro\second}$ and  $\NEP\sim\SI{3E-18}{\watt/\sqrt{\hertz}}$ at temperature $\sim300\,$mK, with a good accomplishment of the theoretical predictions. The response time of our device is faster than a SINIS due to the very reduced heat capacity of graphene compared to metals. The NEP in our device and in the theoretical device of Ref. \cite{golubev2001} are on the same order of magnitude, with a lower value in SIGIS due to the combined effect of a lower base temperature and lower heat dissipation. Another advantage of our device is the reduced heat leakage from the phonons, that is mirrored in low heat transport into the superconducting leads. This prevents the leads overheating, which is a problem present in SINIS systems \cite{courtois2016}. On the other hand, SINIS systems take advantage of well-established fabrication techniques that guarantee high-quality junctions, while techniques for GIS junctions are still in development.

{\it Zero-bias SIGIS bolometers.} Another similar architecture consists of SIGIS devices biased at very low voltage \cite{vora2012,mckitterick2015}. In this case, the electronic refrigeration is absent, and bolometry is performed through the temperature-to-resistance transduction. Theoretically, these devices are predicted to have $\tau\sim \SI{1}{\micro\second}$ and $\NEP\sim\rm2 \times 10^{-19}W/\sqrt{Hz}$ at 100 mK \cite{mckitterick2015}.  In comparison with the theoretical device in Ref. \cite{mckitterick2015}, our device shows a NEP that is one order of magnitude larger but a faster response time. This because the voltage bias increases the junction thermal conductance, thus increasing the noise contribution from the junctions but allowing a faster thermalization. Our device and the zero-bias SIGIS bolometers share the same fabrication issues concerning the quality of the tunnel junctions. At the state of the art, the measured NEP reached in 0V-SIGIS is on the order of $\rm\sim10^{-17}W/\sqrt{Hz}$ \cite{vora2012}.

{\it SNS and SGS Josephson junction bolometers.} Finally, we compare our system with another class of bolometers, based on clean-contacted SNS \cite{govenius2014,govenius2016,kokkoniemi2019} or SGS \cite{lee2019} forming hybrid Josephson junctions. These systems exploit completely different physical phenomena and share with our $V$-biased SIGIS only the materials composing the detector. The transduction involves the temperature-dependence of the junction kinetic inductance or the switching current. 
A recent paper reports a SNS bolometer that, at bath temperature $\rm 25\,mK$, shows a very low NEP $\NEP\sim \rm6\times10^{-20}W/\sqrt{Hz}$ and a quite long response time $\tau=\SI{30}{\micro\second}$. Though, this response time is more than one order of magnitude faster in the class of low noise bolometers \cite{kokkoniemi2019}. Compared to our device, the SNS ultimate experiment shows a longer response time but a better NEP.

A recent pre-print \cite{lee2019} reports a very promising bolometer based on an SGS Josephson junction. The experiment is based on the measurement of the statistic distributions of the switching current (Fulton-Dunkleberger) versus the input power. Then, the NEP is estimated from the width of the distribution, since a larger standard deviation is associated with a larger uncertainty on the power signal measurement. In this way, the Authors estimate a NEP $\NEP\sim\rm7 \times 10^{-19}W/\sqrt{Hz}$, reaching the fundamental limit imposed by the intrinsic thermal fluctuation of the bath temperature at 0.19 K \cite{lee2019}. The SGS-based architecture seems a promising path for further research in the field of low-noise bolometers.

\section{Conclusions and further developments}
\label{sec:summary}
In this paper, we have investigated electron cooling in graphene when tunnel-contacted to form a SIGIS device and its application as a bolometer.

We have studied electron cooling by voltage biasing the junctions, exploiting the same mechanism of a SINIS system. The low electron-phonon coupling in graphene allows having a sensible temperature decrease even for a large area graphene flakes and a high tunnel resistance ($\SI{100}{\micro\meter^2}$, \SI{10}{\kilo\ohm}), differently from a SINIS where a low tunnel resistance is required for adsorbing the larger phonon-heating. 

We have then studied the dynamics of the SIGIS cooler. We obtained the dependence of the thermal relaxation time on temperature and voltage bias and estimated its magnitude ($\tauth\sim 10\,$ns). 

Finally, we have investigated the possibility of employing the cooled SIGIS system for bolometric applications. We found out that electron cooling enhances the responsivity and decreases the noise equivalent power. {Moreover, the small  electron-phonon coupling and the possibility of using high values of tunnel resistance allow reaching low noise equivalent power of the order $\rm 10^{-18}\,  W/\sqrt{Hz}$. At the same time, the cooling mechanism increases the operation speed of the bolometer of more than one order of magnitude. Compared to the unbiased case, this makes the cooled SIGIS a suitable detector for THz communication \cite{kurner2014,nagatsuma2016,ummethala2019} and cosmic microwave background \cite{tarasov2019,inomata2019} applications.}

Further developments for our system could be explored. In particular, many known strategies already employed  to the SINIS coolers/bolometers can be inherited. Among them, suspended graphene can show very interesting cooling characteristics due to the combined refrigeration of electrons and phonons, since in this case the latter are not connected to the substrate thermal bath \cite{koppinen2009,koppinen2009b,koppinen2009c,clark2005}.

\section{Acknowledgments}
The authors acknowledge the European Research Council under the European Unions Seventh Frame-work Programme (FP7/2007-2013)/ERC Grant No. 615187 - COMANCHE, the European Unions Horizon 2020 research and innovation programme under the grant no. 777222 ATTRACT (Project T-CONVERSE), the Horizon research and innovation programme under grant agreement No. 800923 (SUPERTED), the Tuscany Region under the FARFAS 2014 project SCIADRO. M. C. acknowledges support from the Quant-EraNet project SuperTop. The work of F.P. work has been partially supported by the Tuscany Government, POR FSE 2014 -2020, through the INFN-RT2 172800 Project. A.B. acknowledges the CNR-CONICET cooperation program Energy conversion in quantum nanoscale hybrid devices and the Royal Society through the International Exchanges between the UK and Italy (Grant No. IES R3 170054). F.B. acknowledges the European Union’s Horizon 2020 research and innovation program under Grant agreement No. 696656 – Graphene-Core1 and No. 75219 Graphene-Core2. S.R. and A.B. acknowledges the financial support from the project QUANTRA, funded by the Italian Ministry of Foreign Affairs and
International Cooperation.

\end{document}